\documentclass{jfm}
\usepackage{amssymb,amsfonts,amstext,amsmath} 

\usepackage{graphicx}
\usepackage{epstopdf, epsfig}

\usepackage[labelformat=simple, font=small]{subcaption}

\usepackage{url}
\usepackage{aas_macros}
\usepackage[noabbrev]{cleveref}

\crefname{equation}{}{}
\Crefname{equation}{Equation}{Equations}

\newcommand{\bra}[1]{\left( #1 \right)} 
\newcommand{\abs}[1]{\left| #1 \right|} 
\newcommand{\avg}[1]{\left< #1 \right>} 
\newcommand{\cavg}[2]{\left<\left. #1 \right| #2\right>} 

\renewcommand{\v}[1]{\boldsymbol{ #1 }}  

\newcommand{\vx}{\v{x}}

\newcommand{\vk}{\v{k}}
\makeatletter

\shorttitle{Small-scale turbulence and its impact on the pressure field}
\shortauthor{D. G. Vlaykov and M. Wilczek}

\title{On the small-scale structure of turbulence and its impact on the pressure field}

\author{Dimitar G. Vlaykov\aff{1}
 \and  Michael Wilczek\aff{1}  \corresp{\email{michael.wilczek@ds.mpg.de}}}

\affiliation{\aff{1} Max Planck Institute for Dynamics and Self-Organization, Am Fa{\ss}berg 17,37077 G{\"o}ttingen Germany}

\begin{document}

\maketitle

\begin{abstract}
Understanding the small-scale structure of incompressible turbulence and its implications for the non-local pressure field is one of the fundamental challenges in fluid mechanics.
Intense velocity gradient structures tend to cluster on a range of scales which affects the pressure through a Poisson equation.
Here we present a quantitative investigation of the spatial distribution of these structures conditional on their intensity for Taylor-based Reynolds numbers in the range [160, 380].
We find that the correlation length, the second invariant of  the velocity gradient,  is proportional to the Kolmogorov scale.
It also is a good indicator for the spatial localization of intense enstrophy and strain-dominated regions, as well as the separation between them.
We describe and quantify the differences in the two-point statistics of these regions and the impact they have on the non-locality of the pressure field as a function of the intensity of the regions.
Specifically, across the examined range of Reynolds numbers, the pressure in strong rotation-dominated regions is governed by a dissipation-scale neighbourhood.
In strong strain-dominated regions, on the other hand, it is determined primarily by a larger neighbourhood reaching inertial scales.
\end{abstract}


\section{Introduction}

The complexity of turbulent flows arises from two basic properties of the equations of fluid motion, nonlinearity and non-locality, and the intimate connection between them.
The small-scale structures of turbulence are shaped by nonlinear mechanisms, like vortex stretching for example, and are well described by the gradient of the velocity field.
Non-locality, the instantaneous dependence of a fluid element on the state of the entire system, is encoded in  the pressure field in incompressible flows.
In the following study we characterize the spatial structure and length scales associated with the velocity gradients and use them to quantify the effective non-locality of the pressure field.

The statistical and dynamical properties of the velocity gradients have been studied extensively in previous literature.
For a review of experimental results regarding their spatial structure and distribution see e.g. \cite{Wallace2009}.
It is well known that regions of  extreme velocity gradients are organized in small-scale structures.
Traditionally they are identified as vortex tubes and strain sheets \citep{She1990, Douady1991, Jimenez1993}.
These structures are strongly correlated with each other in space (e.g. \citet{Yeung2012, Fiscaletti2014}) and have strongly non-Gaussian probability density functions (PDFs) \citep{Wilczek2009, Yeung2012}.
Their properties are a subject of interest in relation to flame extinction \citep{Sreenivasan1997, Sreenivasan2004}, clustering of particles \citep{Chun2005}, cloud formation \citep{Bodenschatz2010} and mixing \citep{Pumir1994a}.

The pressure field has also been studied in great detail --- see \cite{Cao1999} for a comprehensive study of its spatial structure and statistics;
\cite{Nelkin1994}, \cite{Gotoh1999} and \cite{Vedula1999}  present complementary analyses of the pressure scaling and intermittency properties.
It maintains the incompressibility of the flow and is generally characterized as a large-scale/long-range field (by e.g. \citealt{Monin1975}, pp. 368-377; \citealt{Nelkin1994, Cao1999}).
The pressure PDF is highly non-Gaussian with  an exponential tail for negative pressure fluctuations \citep[see e.g.][]{Fauve1993, Holzer1993, Pumir1994}.
A fundamental understanding of the pressure field has a variety of applications.
From an engineering perspective, very low-pressure regions are interesting because they can trigger cavitation \citep{LaPorta2000}.
The scale of pressure--velocity correlations has a direct impact on the energy decay rate in freely decaying turbulence \citep{Davidson2011}.
In numerical simulations, the pressure non-locality is a major computational bottleneck in real-space-based methods.

Despite the wealth of available literature, 
our understanding of the link between regions of extreme velocity gradient  $\mathsfbi{A} = \nabla \v{u}$
and the pressure $p$ is still incomplete. The incompressible Navier-Stokes equations imply that
 \begin{equation}
 \label{eq: Poisson}
  \nabla^2 p(\vx) = 2 Q_A (\vx).
 \end{equation}
Here, $Q_A$ is the second invariant of $\mathsfbi{A}$, i.e. $Q_A = -\mathrm{Tr} \mathsfbi{A}^2/2$. 
Using Green's method, the solution of the equation is easily seen to be
\begin{equation}
 p(\vx) = \int_{\mathcal{D}} -\frac{   2Q_A (\vx')}{4 \pi |\vx-\vx'|}\mathrm{d}^3 x',
 \label{eq: Greens soln}
\end{equation}
where $\mathcal{D}$ designates the flow domain.
We shall focus on infinite or periodic domains, so we omit boundary effects.
\Cref{eq: Greens soln} shows that at least formally one needs the information about the entire flow in order to compute the pressure at a single point.
This is the essence of the problem of non-locality in incompressible flows and one of the main points of the current study.

In contrast to the formal expression \labelcref{eq: Greens soln}, there have been theoretical, numerical and experimental indications that a good approximation to the statistical properties of the pressure field may be obtained by truncating the integral to a relatively local neighbourhood of the reference point \citep[see e.g.][]{Davidson2011, Constantin2014, Lawson2015}. 
This has implications for the long-range pressure--velocity correlations --- they are suppressed due to the small-scale structure of developed turbulence \citep{Ruelle1990, Ishida2006, Davidson2008, Davidson2011}.
If linked to cancellations between the strain and enstrophy fields, this reduced non-locality effect can lead to reduced intermittency of the pressure spectrum \citep{Nelkin1994}.

Hypotheses for the mechanisms behind the limited non-locality of the pressure have been proposed for several decades. 
There have been comparisons to electrostatic charges \citep{Douady1991, Fauve1993, Pumir1994, Nelkin1994}, charged dipoles of finite size \citep{Gotoh1999} and magnetic dipoles \citep{Ishida2006}. 
Numerous authors have also identified the analogy between \cref{eq: Poisson} and the Debye--H\"uckel theory of shielding in plasmas,
whereby charges of opposite polarity screen each other's effects on the global electric field beyond a critical radius and limit long-range interactions \citep{Batchelor1951, Ruelle1990, Davidson2011}. 

By and large  all these analogies refer to some form of shielding.
In plasmas for example, charge carriers of opposite polarity cluster around each other due to the Coulomb force
and as a result screen each other's large-scale influence on the overall electric field. This leads to limited long-range interactions.
For turbulence the argument proceeds as follows.
The first step is to decompose $Q_A$ into contributions from straining and vortical motions as $Q_A=Q_S + Q_W$.
Here $Q_S = -\mathrm{Tr} \mathsfbi{S}^2/2$ and $Q_W = -\mathrm{Tr} \mathsfbi{W}^2/2$ are the second invariants of the rate-of-strain and rate-of-rotation tensors:  $ \mathsfbi{S} =\bra{\mathsfbi{A} + \mathsfbi{A}^{\mathrm{T}}}/2$ and $\mathsfbi{W}=\bra{\mathsfbi{A} - \mathsfbi{A}^{\mathrm{T}}}/2$  respectively.
The key point is that $Q_S$ is negative semi-definite, while $Q_W$ is positive semi-definite.
In other words, straining (dissipative) and vortical motions have competing effects on the pressure.
As a result, neighbouring regions of strong vorticity and strain effectively shield or cancel each other's effect on the pressure field.
As  already pointed out, in homogeneous and isotropic turbulence the regions of large $Q_S$ and $Q_W$ are in fact organized in tightly entangled and often overlapping small-scale structures. 
This can be inferred from the kinematic relationship between them \citep{Ohkitani1994}.
It is then not too surprising that the resulting pressure fluctuations are much more diffuse and with substantially smaller and less frequent extreme events than could be expected \citep{Nelkin1994}.

\begin{figure}
\centering
	\begin{subfigure}{\textwidth}		
 	\includegraphics[width=\textwidth]{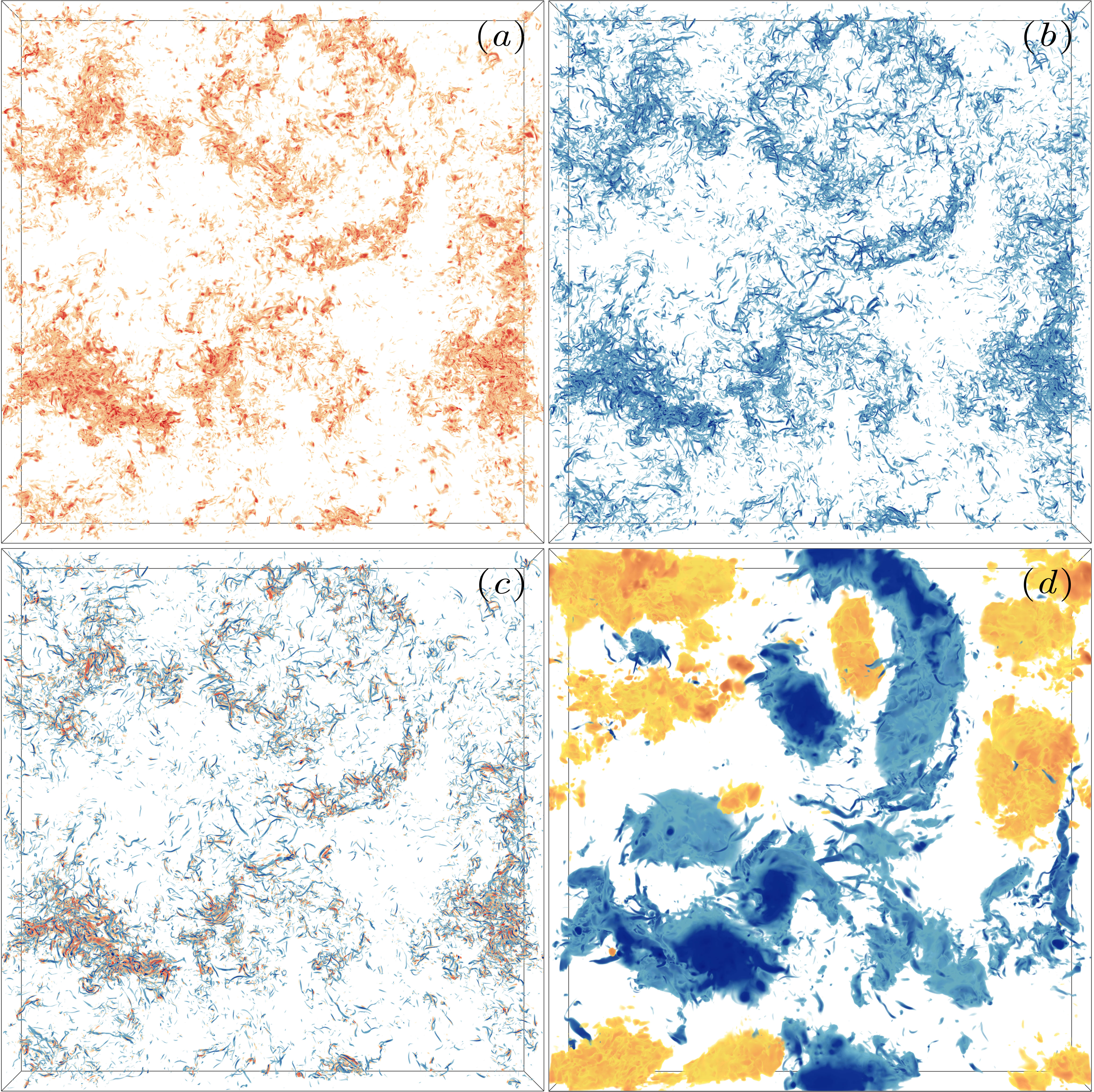}
 	\caption*{\label{fig: visualisations a}}
 	\end{subfigure}

	\begin{subfigure}{0.47\textwidth}		
		\includegraphics[width=\textwidth]{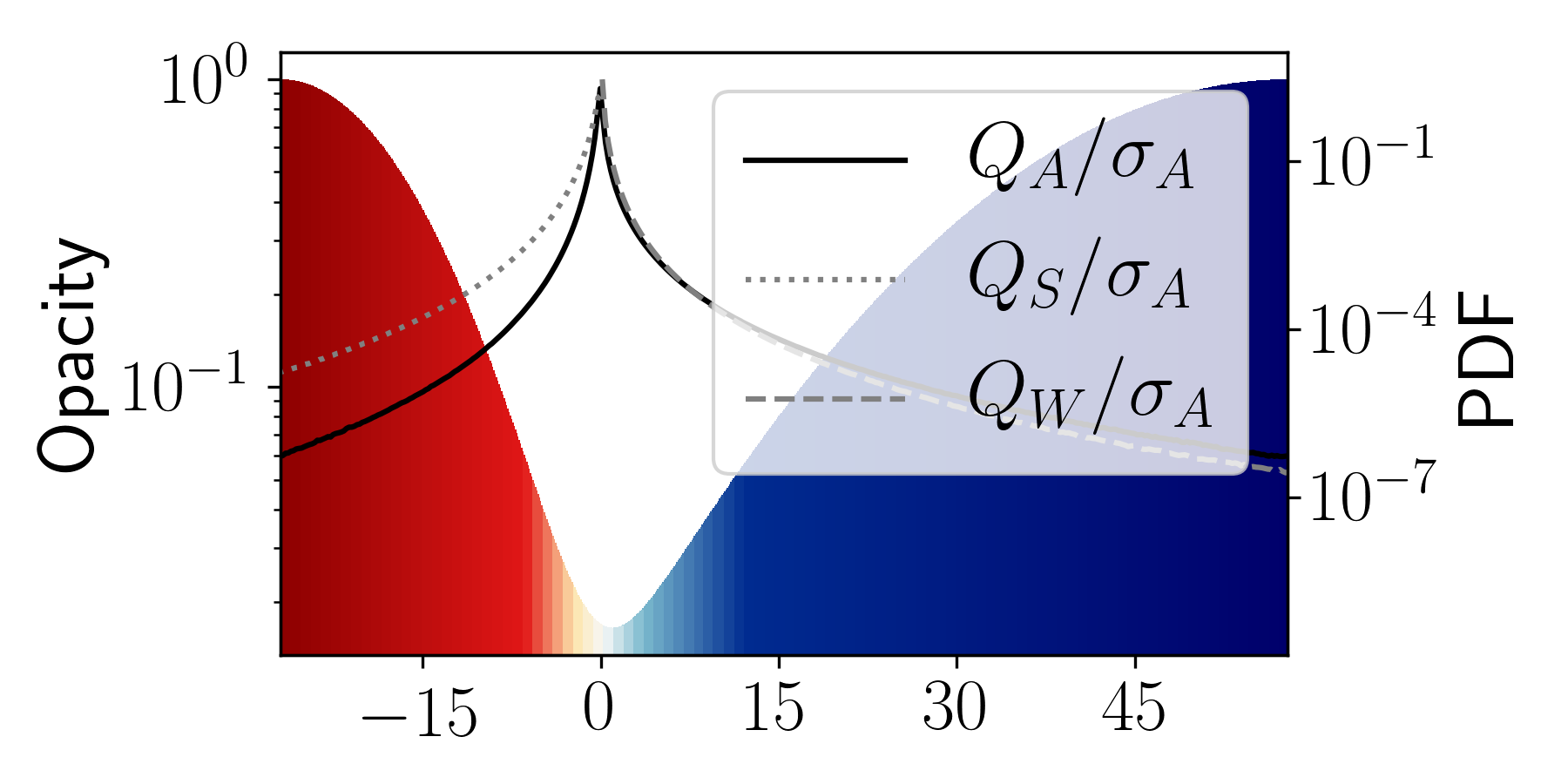}
		\caption*{(\textit{e})\label{fig: visualisations e}}
	\end{subfigure}
	\hfill
	\begin{subfigure}{0.47\textwidth}		
		\includegraphics[width=\textwidth]{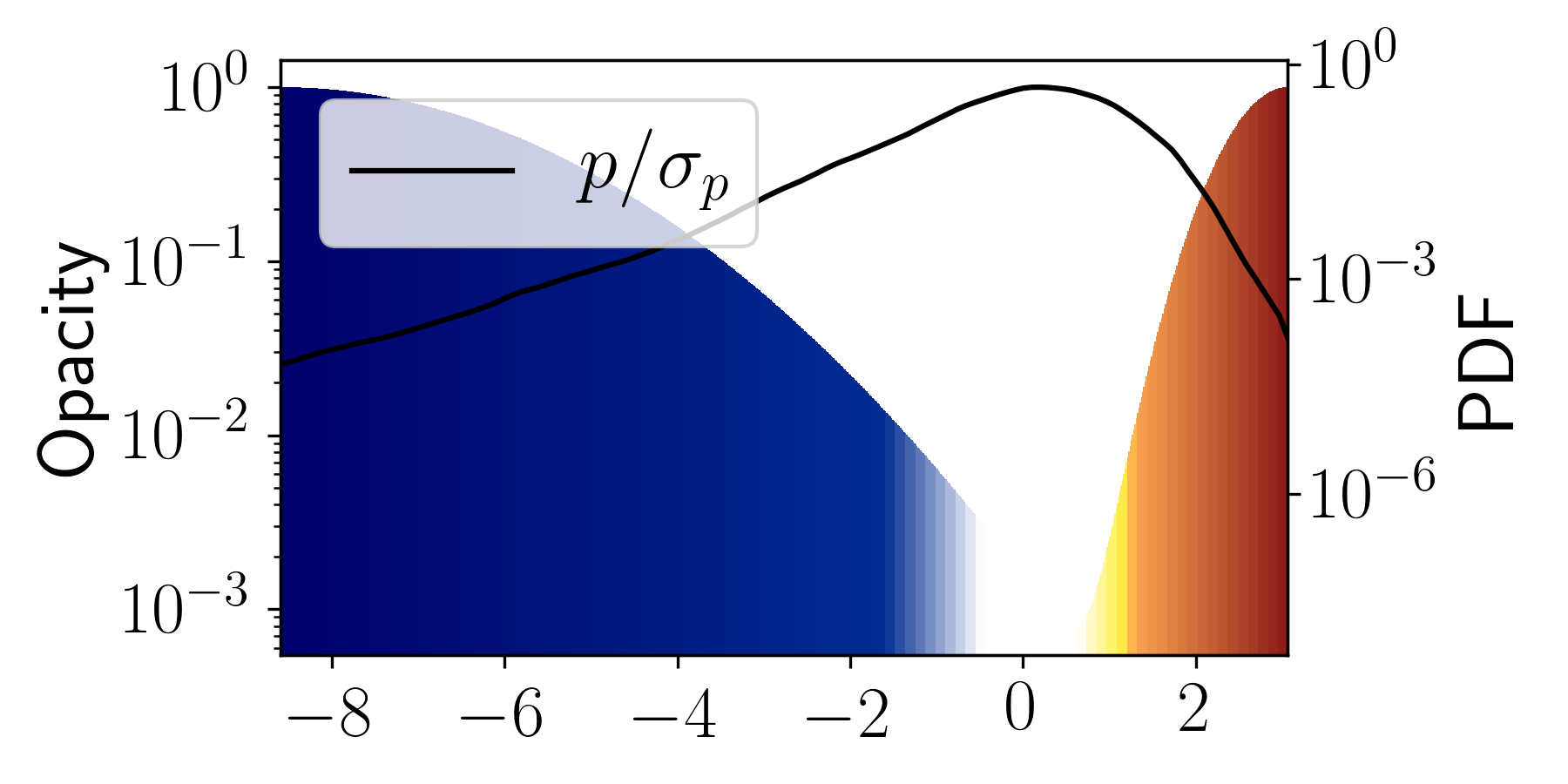}
		\caption*{(\textit{f})\label{fig: visualisations f}}
	\end{subfigure}
	\caption{\small
	  Visualisations of regions of extreme $Q_S$ (\textit{a}), $Q_W$ (\textit{b}), $Q_A$ (\textit{c}) and pressure (\textit{d}), demonstrating the clustering of the the respective components of the velocity gradient on a range of scales and the correlation with extreme pressure fluctuations.
	  The visualisations are from numerically simulated data with $\Rey_\lambda =375 $, for more details about the simulation see \cref{sec: numerics} and \cref{table: data}.
	  The volume shown is $2576 \eta$ on the side and $258 \eta$ deep, corresponding to a layer of the simulated domain with aspect ratio 1/10.  
	  The bottom two panels show the colour-opacity maps (shaded regions) and PDFs (curves) for $Q_S$, $Q_W$  and $Q_A$  in (\textit{e}) and for the pressure in (\textit{f}).
	  The visualisations have been produced with the open source application VAPOR \citep{Clyne2005, Clyne2007} using colour maps from \cite{colorbrew}. 
	  \label{fig: visualisations}}
\end{figure}

A visualization of the extreme parts of the $Q_S$, $Q_W$, $Q_A$ and $p$ fields from a simulation of homogeneous and isotropic turbulence (see \cref{fig: visualisations}) lends qualitative support to the shielding paradigm.
The low-intensity parts of the fields are rendered transparent using the shown colour-opacity maps. 
The highlighted structures are at ten or more standard deviations for the velocity gradient invariants and at two or more standard deviations for the pressure,
as shown by the corresponding PDFs in panels (\textit{e}) and (\textit{f}).
The small-scale structure of intense vorticity and strain regions, as well as their strong correlation, is evident in panels (\textit{a}) and (\textit{b}).
In panel (\textit{c}) the tight entanglement between regions of positive and negative $Q_A$ can be seen, i.e. regions dominated by the vorticity or the strain.
The $Q_A$ structures are characteristically smaller, more spatially localized and even more entangled than the corresponding $Q_S$ and $Q_W$ regions.
The intense pressure structures shown in panel (\textit{d}) are a combination of large diffuse regions and tight spatially localized low-pressure structures.
The former correspond to large voids of low-intensity $Q_A$ field, while the latter correspond to the sharply localized extreme $Q_A$ structures, i.e. to the vortex tubes and strain sheets.

To the best of our knowledge no direct quantitative investigations of the shielding hypotheses have been performed to date. 
This motivates the following study of homogeneous and isotropic turbulence. 
Its purpose is twofold.
Firstly, it provides quantitative measures for the characteristic sizes and separations between velocity gradient structures of different types and intensity.
Secondly, it quantifies the impact of the velocity-gradient structures on the non-locality of the pressure field.
The focus will fall primarily on the intense small-scale velocity-gradient  and pressure structures, which contribute to higher-order statistics.
We will show to what extent for homogeneous and isotropic turbulence, the pressure at extreme events depends on the properties of the local structures and the far-field contributions.
We shall establish the dependence of this behaviour on the intensity of the local structure (as measured by the velocity gradient fields) and on the Reynolds number. 
The volume-filling large-scale low-intensity regions shall not be addressed in much detail.
Their properties are well captured by low-order statistics which have been discussed by e.g. \cite{Cao1999,Vedula1999}.

We begin in \cref{sec: background} with some basic considerations of the properties of $Q_A$ and their impact on the pressure.
In \cref{sec: numerics} we describe the numerical simulations used in the analysis.
This is followed by an analysis of  the statistical properties of $Q_A$  and their dependence on the local conditions and the Reynolds number
with the help of highly resolved direct numerical simulations (DNS) of statistically stationary, homogeneous and isotropic turbulence (\cref{sec: vel gdt}).
The role of the Poisson weighting kernel is identified and the behaviour of the contributions to pressure field is analysed in \cref{sec: pressure}, before final conclusions are drawn.

\section{Background}
\label{sec: background}

To begin with consider the effect of a turbulent region on the pressure at a reference point $\vx$. 
Since the Poisson kernel is spherically symmetric, we can isolate its effect by decomposing \cref{eq: Greens soln} into radial and angular integrals. 
For brevity we denote the spatial angle average of a field with an overbar.
For example,

\begin{align} 
  \overline{Q_A}(\vx, r) &= \frac{1}{4 \pi} \int_{\Omega_{\vx}(r)} Q_A(\vx') \mathrm{d} \Omega,
  \label{eq: angl ave def}
 \end{align}
where $\Omega_{\vx}(r)$ is a thin spherical shell centred on $\vx$ with a radius $r$ and $\mathrm{d} \Omega$ is the surface element on the unit sphere. 
Then the pressure at $\vx$ is given by 
\begin{align}
  p(\vx) &= \int \limits_{0}^{\infty} \overline{q_A}(\vx, r)  \mathrm{d} r = - \int \limits_{0}^{\infty} 2  r \overline{Q_A}(\vx, r)  \mathrm{d} r ,
  \label{eq: rad p check}
 \end{align}
where 
$\overline{q_A}(\vx, r)$ is shorthand for the contributions to the pressure at $\vx$ from $\Omega_{\vx}(r)$,
 \begin{align}
   \overline{q_A}(\vx, r) = -2 r\overline{Q_A }(\vx, r).
   \label{eq: qA def}
 \end{align}

Thus, the question of the locality of the pressure becomes one about the properties of $\overline{q_A}(\vx, r)$ or consequently about the properties of $\overline{Q_A }(\vx,r)$.

To put a first bound on the behaviour of $\overline{Q_A}(\vx , r)$, consider the large-$r$ limit.
Since, $Q_A$ is statistically homogeneous, it is immediate that
\begin{align}
\overline{Q_A}(\vx, r) \rightarrow \avg{Q_A} \textrm{ for } r \gg L_{A},
\label{eq: omega limit}
\end{align}
where $L_{A}$ is the correlation length of $Q_A$ and $\avg{Q_A}$ is the volume average of $Q_A$ (or under ergodicity its ensemble average). 
The value $\avg{Q_A}$ is approached whenever an average of $Q_A$ is taken over a sufficient number of uncorrelated points, i.e. points separated by distances much larger than $L_{A}$.
Here,  we define
\begin{equation}
L_{A} =  \int_0^{\infty} \abs{\mathcal{C}_{A} (r)} \mathrm{d} r,
\label{eq: int auto length scale}
\end{equation}
where the $C_{Q_A}(r)$ is the auto-correlation function of $Q_A$, i.e.

\begin{equation}
 \mathcal{C}_{A} (r) = \frac{1}{\sigma_A^2 }\bra{\avg{Q_A(\vx) Q_A (\vx')} - \avg{Q_A}^2},
\end{equation}
with $r = \abs{\vx-\vx'}$ and $\sigma_A$ -- the standard deviation of $Q_A$.
The absolute value in \cref{eq: int auto length scale} is taken to allow for changes in the sign of the correlation function. 
In this way correlated and anti-correlated regions are treated equally.

In a periodic or infinitely extended domain $\avg{Q_A}=0$ and therefore $\overline{Q_A }(\vx, r)$ decays to zero for $r \gg L_{A}$.
In conjunction with \cref{eq: rad p check} this suggests that,if the decay is fast enough, the dominant contributions to $p(\vx)$ will come from correlated regions in space not much bigger than $L_{A}$.
The question of locality is then transferred to the size and spatial distribution of these regions. 

For future reference we shall denote with $\mathcal{C}_{SW}$ the cross-correlation between $Q_W$ and $Q_S$ and with $L_{SW}$ the associated correlation length.
They are given respectively by 

\begin{equation}
 \mathcal{C}_{SW}(r) = \frac{1}{\sigma_{S}\sigma_{W}}\bra{\avg{Q_S(\vx) Q_W(\vx')}   - \avg{Q_S} \avg{Q_W}},
\end{equation}
and 
\begin{equation}
L_{SW} =  \int_0^{\infty} \abs{\mathcal{C}_{SW}(r)}\mathrm{d} r.
\label{eq: int cross length scale}
\end{equation}

\section{Numerical details}
\label{sec: numerics}

To proceed beyond the basic analytical consideration above, we use direct numerical simulations of homogeneous and isotropic turbulence.
The simulation data are generated by a pseudo-spectral solver \citep{Lalescu2018} in a three-dimensional periodic domain of length $L=2 \pi$ (code units). 
Full details about the simulation parameters can be found in \cref{table: data}.
Statistical stationarity is achieved through band-passed large-scale Lundgren forcing \citep{Lundgren2003, Rosales2005} 
for the majority of the simulations.
To test the robustness of the results to different types of large-scale forcing,
in simulation 1 the overall forcing amplitude is also adjusted in time in a way which ensures fixed energy injection rate.
The chosen forcing wavenumber range reduces the effect of the periodic boundary conditions by allowing the velocity auto-correlation to decay within the simulation domain:
it accommodates  between seven and eleven velocity correlation lengths.
The de-aliasing of the small scales is performed using a smooth exponential filter \citep{Hou2007}.
A third-order memory-saving Runge-Kutta scheme is used for the time integration \citep{Shu1988}.
We investigate Taylor-based Reynolds numbers in the range $  160- 380$,
using small-scale resolutions in the range   $2 \lesssim k_{\textrm{max}} \eta \lesssim 4$.
Here $\eta$ is the Kolmogorov scale and $k_\textrm{max}$ is the maximum fully resolved wavenumber.
The larger resolution permits a closer look at the details on the small scales, as systematically studied by e.g. \cite{Schumacher2007}.
It also allows us to assess the uncertainty of the presented results with respect to resolution effects.
To this end simulations 6 and 7 are designed to have approximately the same Reynolds numbers.

A number of instantaneous snapshots are taken from each simulation and treated as independent realizations of a turbulent field with the same parameters in order to achieve statistical convergence.
They are separated by at least a third of the integral time $T_\textrm{int}$,
so that temporal correlations on the small scales characterizing the velocity gradient have decayed sufficiently.

\begin{table}
\centering 
\makebox[\textwidth]{
 \begin{tabular}{|c|c|c|c|c|c|c|c|c|c|c|c|c|}
 
 \hline
  \# & $\Rey_\lambda$ & $k_{\textrm{max}} \eta$ & $\Delta/\eta$ & $N^3$ &  $L_\textrm{int}/\eta$ & $L/L_\textrm{int} $ & $T_\textrm{int}/\tau_\eta$ & $U_\textrm{int}$ & $\avg{Q_W}$  & $N_T$ & $\delta T/T_\textrm{int}$& $k_f$
  \\
   \hline                                                                       
   1 & 375 & 2.01 & 1.25 & $2048^3$& 368 & 7  & 37.4 & 0.97 &406  & 14 & 4.3 & [1.4, 2.3]
   \\                                                                      
   2 & 317 & 1.91 & 1.32 & $2048^3$& 302 & 9  & 33.3 & 1.01 &573  & 10 & 3.9 & [1.5, 3]
   \\                                                                      
   3 & 279 & 1.86 & 1.35 & $2048^3$& 255 & 11 & 30.0 & 0.91 &556  & 13 & 7.1 & [2, 4]
   \\                                                                         
   4 & 223 & 1.91 & 1.32 & $1536^3$& 182 & 11 & 24.0 & 0.87 &344  & 14 & 6.7 & [2, 4]
   \\                                                                      
   5 & 199 & 1.97 & 1.28 & $1344^3$& 157 & 11 & 21.9 & 0.84 &255  & 19 & 60.0& [2, 4]
   \\                                                                      
   6 & 168 & 3.82 & 0.66 & $2048^3$& 124 & 11 & 18.8 & 0.87 &201  & 11 & 4.5 & [2, 4]
   \\                                                                      
   7 & 163 & 1.97 & 1.28 & $1024^3$& 119 & 11 & 18.3 & 0.83 &177  & 20 & 20.3& [2, 4]
   \\
   \hline
 \end{tabular}
 }
 \caption{\label{table: data} \small Details of the simulation suite describing the main parameters normalized in several ways.
 The following notation is used: 
 Taylor-based Reynolds number $\Rey_\lambda$,
 maximum resolved wavenumber $k_\textrm{max}$ in units of the Kolmogorov length scale $\eta$,
 grid-spacing in real space $\Delta$,
 number of cells in the cubic simulation domain $N^3$,
 integral length scale of the longitudinal velocity correlation $L_\textrm{int}$,
 size of simulation domain $L=2\pi$ in code units,
 ratio of the integral time scale $T_\mathrm{int}= L_\mathrm{int}/U_\mathrm{int}$ to
 the Kolmogorov time scale $\tau_\eta $, 
 integral velocity scale $U_\mathrm{int} = \sqrt{\avg{\v{u}^2}/3}$ in code units, 
 average of $Q_W$ over the simulation domain and considered snapshots $\avg{Q_W}$ in code units, 
 number of analysed snapshots $N_T$ in a time period $\delta T$ and forcing wavenumber range $k_f$.
 }
\end{table}

\section{Analysis of velocity gradients of turbulent flow}
\label{sec: vel gdt}
\subsection{Unconditional statistics}
\label{sec: uncond vel gdt}
As a first step towards quantifying the length scales of the velocity gradient
we compute the two-point correlations of $Q_A$, $Q_S$ and $Q_W$.
Note that these are effectively fourth-order statistics of the velocity gradient $\mathsfbi{A}$.
It should be pointed out that intense vorticity is typically organized into quasi one-dimensional tight vortex tubes,
while intense strain tends to be rather two-dimensional \citep{Moisy2004}.
So, inevitably, the scalar two-point correlations 
represent these different geometries in a highly condensed and simplified fashion. 
Nevertheless, the
auto-correlations indicate typical average length scales of the fields, 
while the cross-correlations indicate a typical average separation between them.

\begin{figure}
\centering
	\begin{subfigure}{0.5\textwidth}
		\includegraphics[width=\textwidth]{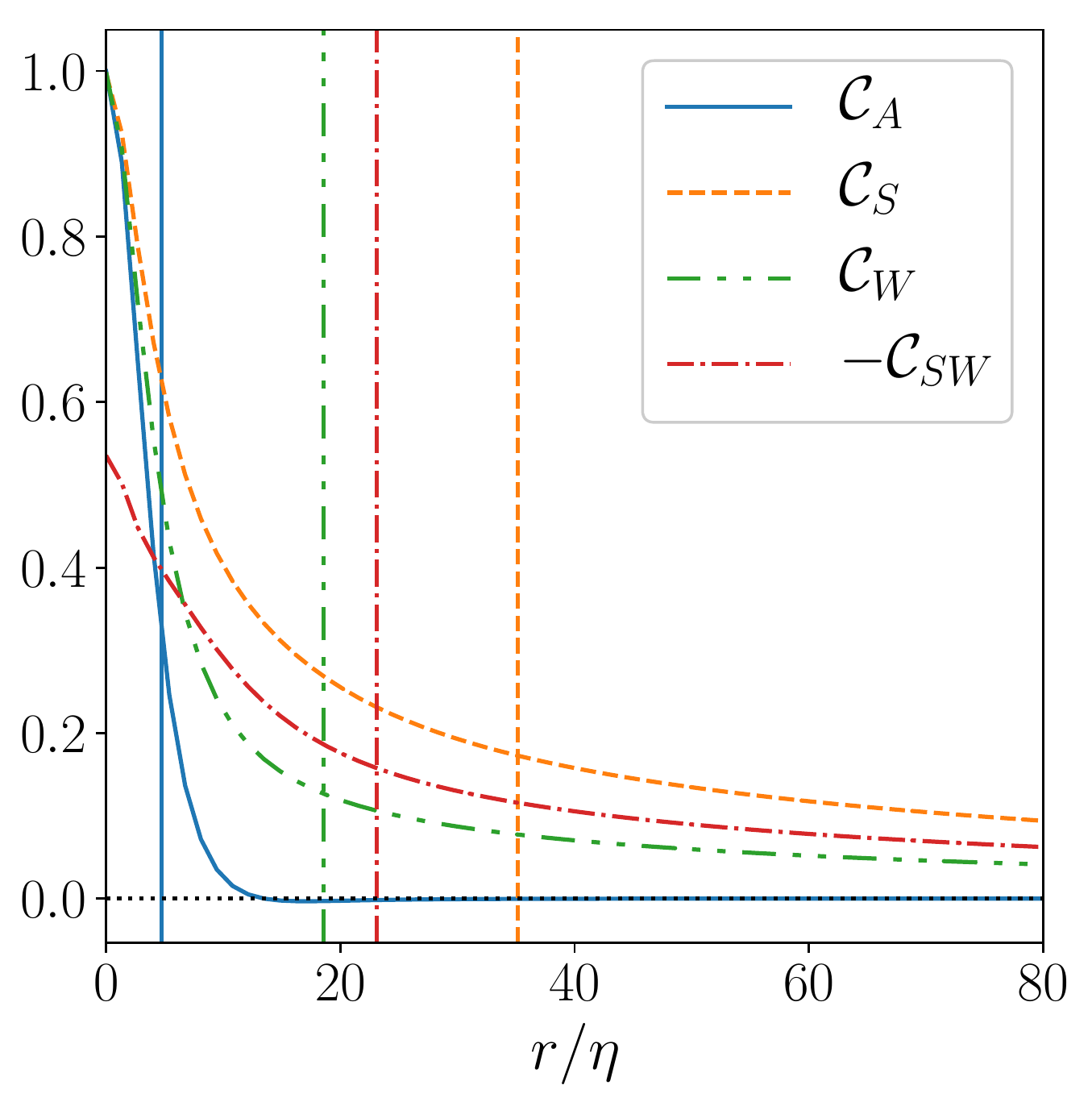}		
		\caption{\label{fig: SW corr a}}
	\end{subfigure}%
	\begin{subfigure}{0.5\textwidth}
		\includegraphics[width=\textwidth]{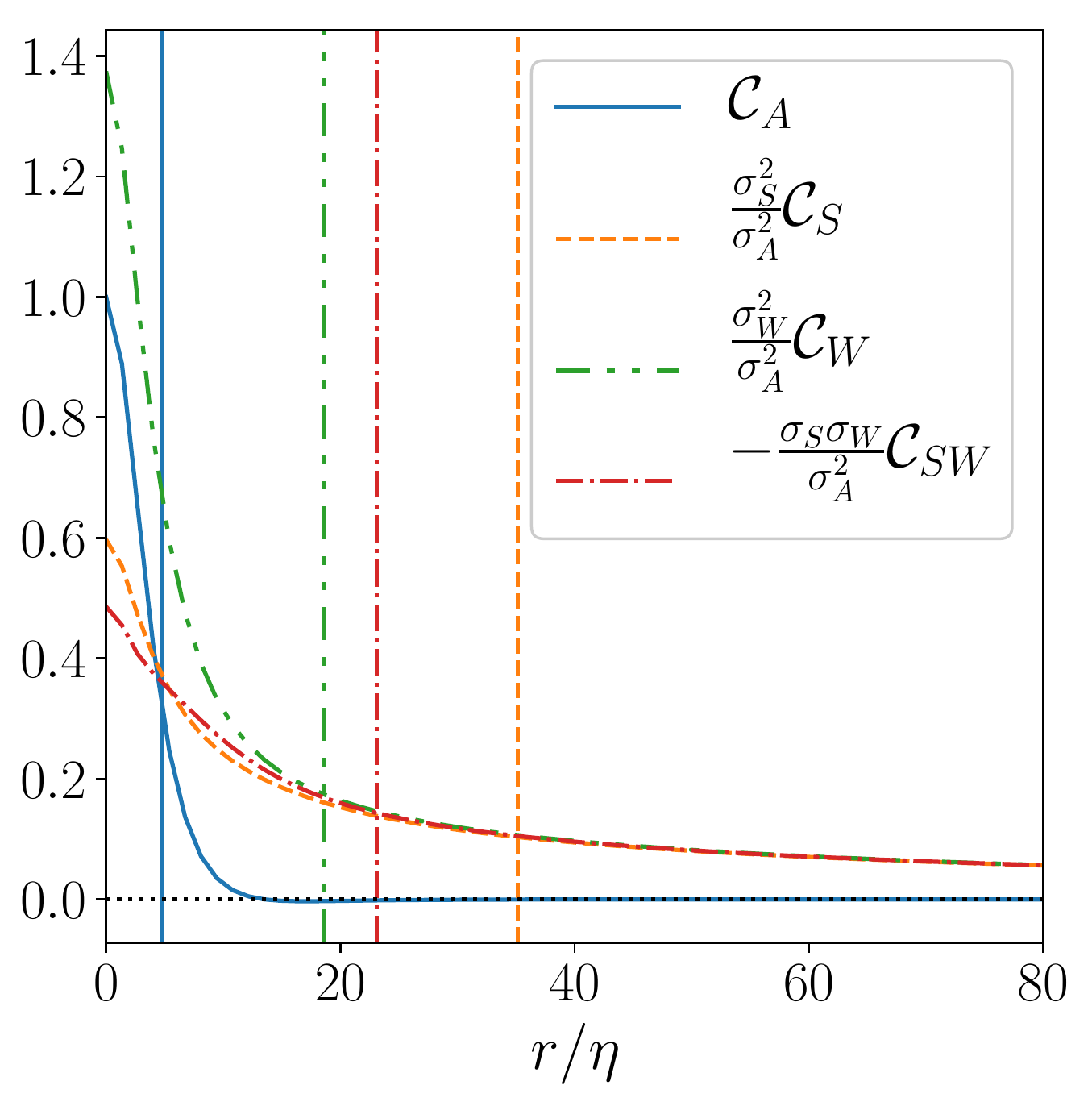}
		\caption{ \label{fig: SW corr b}}
	\end{subfigure}	
	\caption{
        \small
        Spatial correlations of the velocity gradient invariants for $Re_\lambda = 279 $.
        Only the central portion of the correlations is shown to emphasize their different small-scale properties.
	(\textit{a}) Auto-correlation function of $Q_A$ and its constituents --- the auto-correlations of $Q_S$ and $Q_W$ and their cross-correlation). 
	(\textit{b}) The correlations from (\textit{a}) rescaled by the respective standard deviations emphasizing the almost perfect agreement of  the components of \cref{eq: qa var decomp} beyond $\approx 20 \eta$.
	The vertical lines in both panels mark the position of the respective correlation lengths as defined by \cref{eq: int auto length scale,eq: int cross length scale}.
	The full simulation domain, which extends to $2805\eta$, is used in the calculation of these length scales.
	\label{fig: SW corr}
	}
\end{figure}

As an example \cref{fig: SW corr a} shows all auto-correlations as well as the cross-correlation between $Q_S$ and $Q_W$ for the flow with $Re_\lambda = 279$.
Only the region up to $80 \eta$ is shown in order to emphasize the different decaying profiles.
The full simulation extends up to $\approx 2805 \eta$ (see \cref{table: data}).
The correlation lengths associated with the correlation functions are computed using \cref{eq: int auto length scale,eq: int cross length scale} and their analogues for $Q_S$ and $Q_W$. 

Clearly, $Q_A$ is much more shortly correlated than $Q_W$ which in turn is more shortly correlated than $Q_S$. 
The short correlation length scale of $Q_A$ is due to cancellations between $Q_S$ and $Q_W$, 
when enstrophy and strain structures overlap. This can be seen in \cref{fig: visualisations} and is supported by the decomposition
\begin{align}
 \mathcal{C}_{A}(r) =  \bra{\frac{\sigma_S}{\sigma_A}}^2\,\mathcal{C}_{S}(r)+\bra{\frac{\sigma_W}{\sigma_A}}^2\,\mathcal{C}_{W}(r)+ 2\, \frac{\sigma_S \sigma_W}{\sigma_A^2}\,\mathcal{C}_{SW}(r).
 \label{eq: qa var decomp}
\end{align}
\Cref{fig: SW corr b} illustrates this decomposition for the flow with $Re_\lambda=279$.
Note that by definition $\mathcal{C}_{SW }(r)\leq 0 $, while  $\mathcal{C}_{W}(r) \geq 0  $ and $\mathcal{C}_{S} (r) \geq 0$.
So, the auto-correlations of ${Q_W}$ and ${Q_S}$ enhance the auto-correlation of $Q_A$ and the cross-correlation between $Q_S$ and $Q_W$ decreases it.
At the very small scales $\mathcal{C}_{A}$ is dominated by enstrophy correlations.
The lack of correlations beyond approx. $ 20 \eta$ is the result of a fine balance between the length scales of enstrophy and strain structures on the one hand, and the separation between them on the other. 
A crucial component which maintains the balance is the weighting by the respective standard deviations.
In the end the appreciable $Q_A$ correlations are the residue from small-scale enstrophy and strain structures.

\begin{figure}
\centering 
\begin{subfigure}{\textwidth}
 \includegraphics[width=\textwidth]{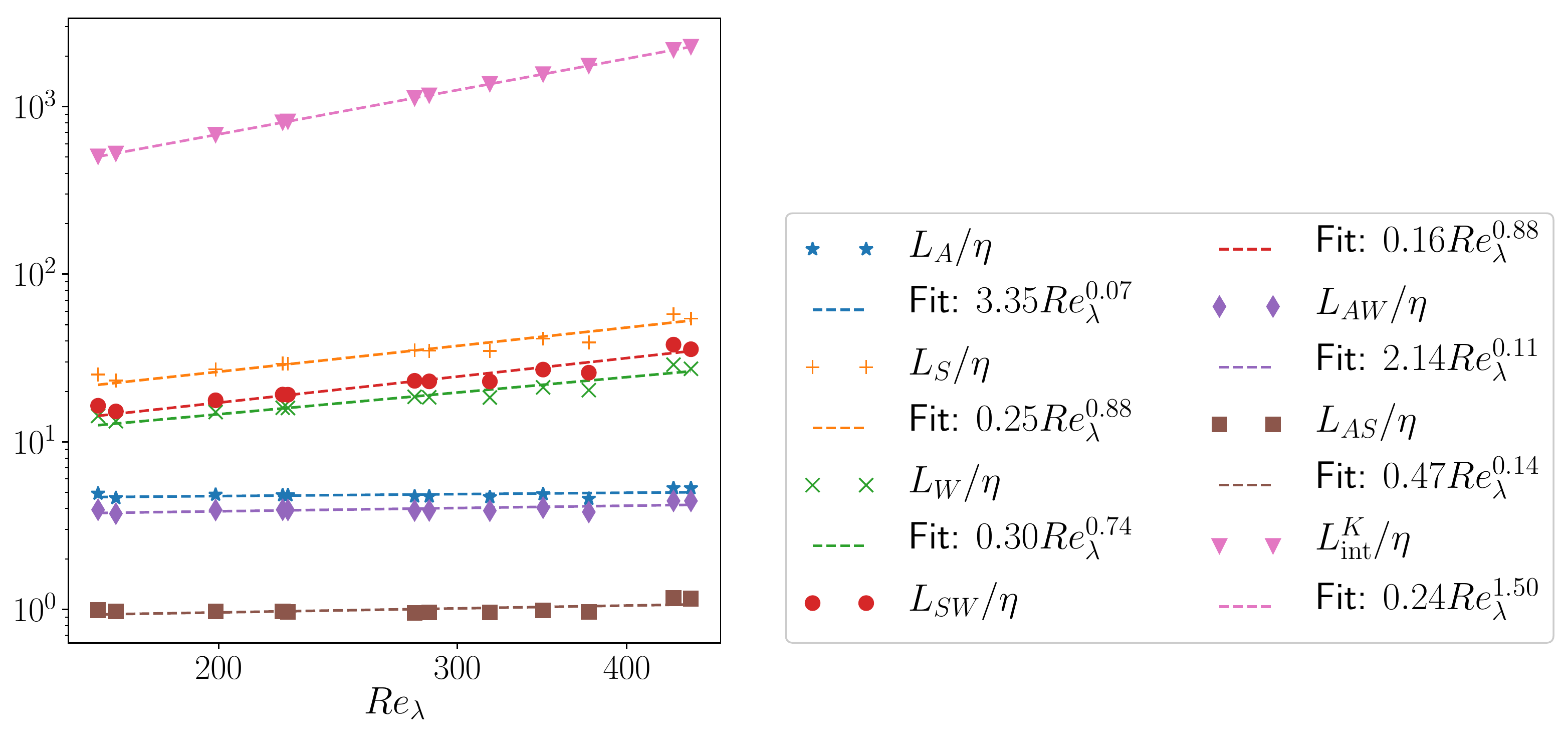}
\end{subfigure}
 \caption{
 \label{fig: length scale Re dependence}
 \small 
 Dependence of the correlation lengths of velocity gradient correlations on $\Rey_\lambda$. 
 The velocity integral length scale $L_\mathrm{int}^K$ is plotted for reference as well.
 All length scales are normalized by the Kolmogorov scale.
 The symbols show the DNS results, the dashed lines --- the corresponding best power law fits (in a least square sense).
 The scaling of $L_\mathrm{int}^K$ agrees with the theoretical expectation.
 Of note is the extremely weak dependence of $L_{A}$ on $\Rey_\lambda$.
 }
\end{figure}

All simulations show the same qualitative behaviour. 
To quantify it, we plot the $\Rey_\lambda$ dependence of the associated correlation lengths  in \cref{fig: length scale Re dependence}.
The length scales are normalized by the Kolmogorov scale.
As a reference we show the scaling of the velocity integral length scale  $L_\mathrm{int}^K = K^{3/2}/\varepsilon$, which agrees with the theoretical scaling of $\Rey_\lambda^{3/2}$ \cite[p.200]{Pope2000}.
Here $K$ is the kinetic energy of the velocity fluctuations and $\varepsilon = 4 \nu \avg{Q_S}$ is the mean dissipation rate.
As can be seen, the $\Rey_\lambda$ dependence of all length scales is fitted quite well by power laws. 
Remarkably, even though $L_{S}$, $L_{W}$ and $L_{SW}$ do not scale with the Kolmogorov scale (they grow with $\eta$), $L_{A}$ does.
In fact, a direct linear fit gives $L_{A}/\eta = 4.5 +1.3 \times10^{-3} \Rey_\lambda $ within the range of investigated Reynolds numbers, with error comparable to the one of the power-law fit.
This indicates that structures of $Q_A$ are characterized well on average by the $4.5\eta$ scale,
in agreement with \cite{Lawson2015}.

\subsection{Conditional statistics}
\label{sec: cond vel gdt}
The considerations above describe coarsely a characteristic  length scale of $Q_A$, but disregard completely the structural differences between regions of small and large intensity or positive and negative sign. 
For instance, regions of smaller intensity tend to be more diffuse and less structured \citep{Moisy2004}.
In order to disentangle the properties of the different regions we continue by examining the $Q_A$, $Q_S$ and $Q_W$ fields in the neighbourhood of strain- and enstrophy-dominated regions of different intensities using two-point conditional statistics.
For simplicity, we focus on the radial properties of these regions, averaging out the directional information. 
This also facilitates the quantification of the non-locality of the pressure in these regions discussed in \cref{sec: pressure}.

\begin{figure}
   \includegraphics[width=\textwidth]{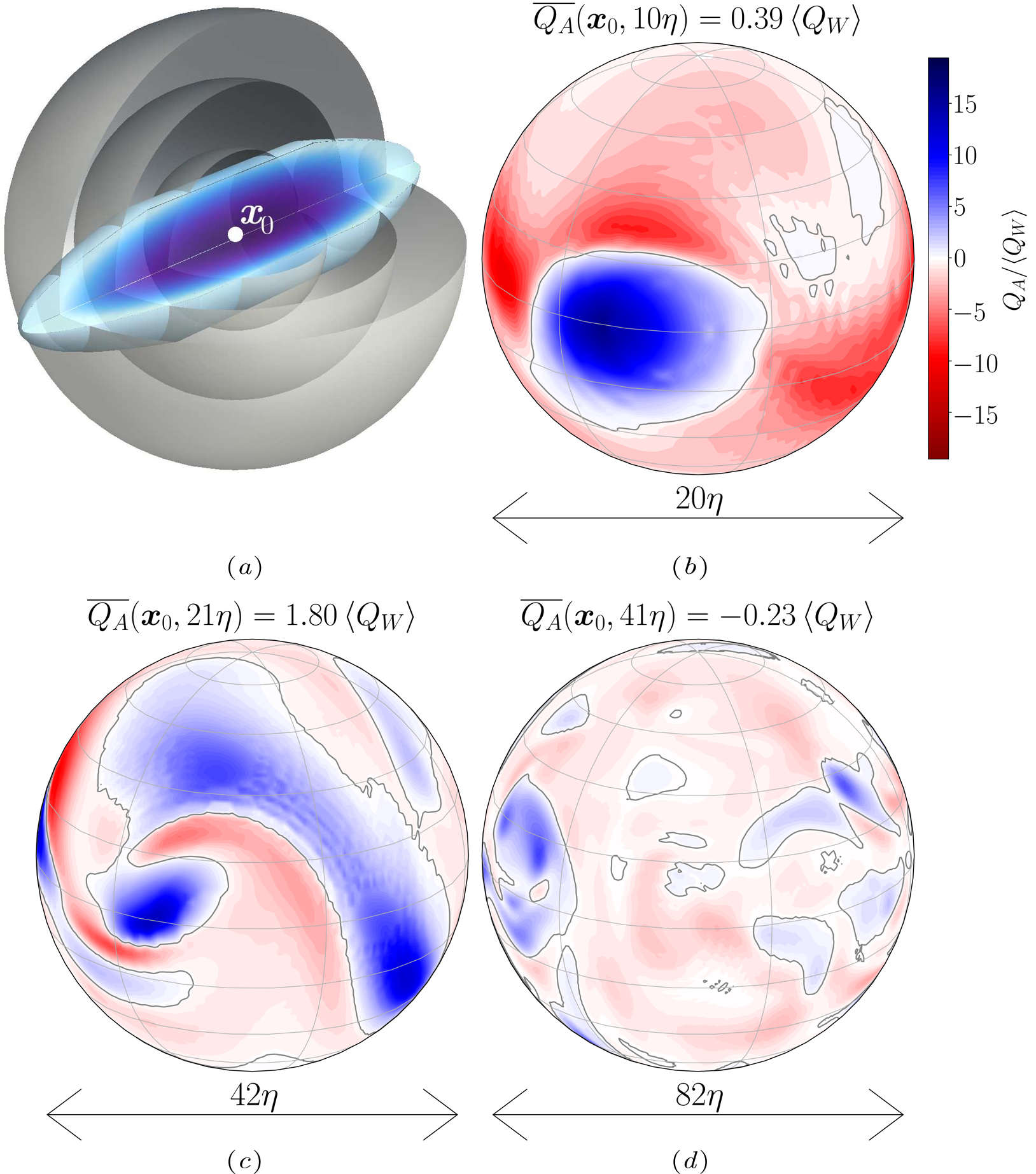}
 \caption{
 \label{fig: turbulent sky}
 \small
  Illustration of the distribution of $Q_A$ on concentric shells of increasing radius centred on a vortex tube.
  The data is taken from the simulation with $\Rey_\lambda = 168$ (see \cref{table: data}).
  The central point has $Q_A(\vx_0)\approx 45 \avg{Q_W}$.
  (\textit{a}) A sketch of the intersection of the vortex tube with spherical shells. 
  (\textit{b-d})
  Colour maps of the $Q_A$ field at distances $10\eta$, $21\eta$ and $41\eta$ from the central point, respectively.  
  The grey contour identifies the $Q_A=0 $ level.
  The average over the shell, $\overline{Q_A}(\vx_0,r)$  is given for each panel. 
  The ringing in some of the panels is an artefact from the cubic interpolation of the Cartesian-grid data to polar co-ordinates and does not represent inaccuracy in the computational data itself.   
  }
\end{figure}

In order to understand the structure of these fields and interpret their statistics, it is illustrative to begin with a simpler example.
Consider how the $Q_A$ field is organized around a point in a turbulent flow dominated by enstrophy.
To illustrate the kinematic relationship between strain and enstrophy we show this in \cref{app: analytic example} for an isolated Townsend's eddy.
Here, instead we consider the structure of $Q_A$ on spherical shells of increasing radius around a reference point with $Q_A (\vx_0) \approx 45 \avg{Q_W}$ in a turbulent flow with $\Rey_\lambda = 168$.
The reference point belongs to a vortex tube, as sketched in \cref{fig: turbulent sky}(\textit{a}).
So, for distances smaller than the vortex tube's radius, in this case for $r\lesssim 3\eta$, $Q_A$ is large and positive.
The spatial distribution of $Q_A$ at larger distances is shown in figures \labelcref{fig: turbulent sky}(\textit{b}) - \labelcref{fig: turbulent sky}(\textit{d}).
\Cref{fig: turbulent sky}(\textit{b}) shows that the vortex tube is embedded in a strongly strain-dominated neighbourhood.
As the radius increases further, the visualization surface grows beyond the vortex tube's length and 
typical turbulence contours emerge. Thus figures \labelcref{fig: turbulent sky}(\textit{c}) and \labelcref{fig: turbulent sky}(\textit{d}) show the footprints of neighbouring entangled vortex tubes and strain structures.
It is evident in all panels that regions of enhanced enstrophy are neighbouring regions of enhanced strain (as dictated by the kinematic relationship between them \citep{Ohkitani1994}) demonstrating the correlation between the two fields on the level of individual structures.
This leads to cancellations when computing e.g. the average of $Q_A$ at a given distance from the reference point, i.e. $\overline{Q_A}(\vx, r)$. It decays quickly from the value at the reference point $45 \avg{Q_W}$ to its overall mean of zero, albeit with some fluctuations. 
In particular, the decay is already significant at $\approx 10 \eta$, where the structure of the vortex tube is still clearly identifiable.

In the following we test the statistical relevance of these observations by analysing the  conditional averages of $\overline{Q_A}(\vx,r)$ around reference points $\vx$ with different values of $Q_A$, i.e. $\cavg{\overline{Q_A}(r)}{Q_A}$.
To isolate the effect of cancellations between vorticity and strain structures, we also calculate the corresponding conditional averages of $\overline{Q_S}(\vx,r)$ and $\overline{Q_W}(\vx,r)$.
The angle-averaged fields arise naturally when considering the generation of the pressure field (see \cref{eq: rad p check}). 
In their own right, they serve to characterize the isotropic component of the velocity gradient structures.
(See e.g. \citet{Biferale2005} for a review on the effects of anisotropy on turbulent statistics.)
We condition on the local value of $Q_A$ to separate properties of enstrophy- and strain-dominated reference points.
Note, however, that this is not equivalent to separating between regions of strong enstrophy and shear because of the strong overlap between the two (as observed experimentally by e.g. \citet{Fiscaletti2014} for a jet flow  and \citet{Worth2011} in a mixing tank).

Conditional averages are an established tool to describe turbulent structures.
One-point conditional averages have been used to associate low-pressure regions with intense enstrophy \citep{Cao1999} and with intense enstrophy and strain \citep{Yeung2012}.
In the context of velocity gradients, two-point conditional averaging of $Q_A$  has been used to study numerically and experimentally the dominant flow structures for different types of local topologies \citep{Lawson2015} and in the context of a stochastic estimation model \citep{Wilczek2014b}.
In the former it is found that intense strain and vorticity structures of various geometries are usually in the neighbourhood of one another in agreement with previous literature. 
Additionally, geometrical conditioning has been used to establish a dominant structure for small-scale turbulence \citep{Elsinga2010} consisting of a shear layer coincident with a pair of stretched co-rotating vortices, the core of which scales with $\eta$. 
The following analysis provides complementary information to these studies.
Firstly, we use conditional averaging to characterize structures of different types and intensity.
Secondly, we quantify the scale of these structures along with their typical separation.

\begin{figure}
 \begin{subfigure}{0.33\textwidth}
 	 \includegraphics[width=\textwidth]{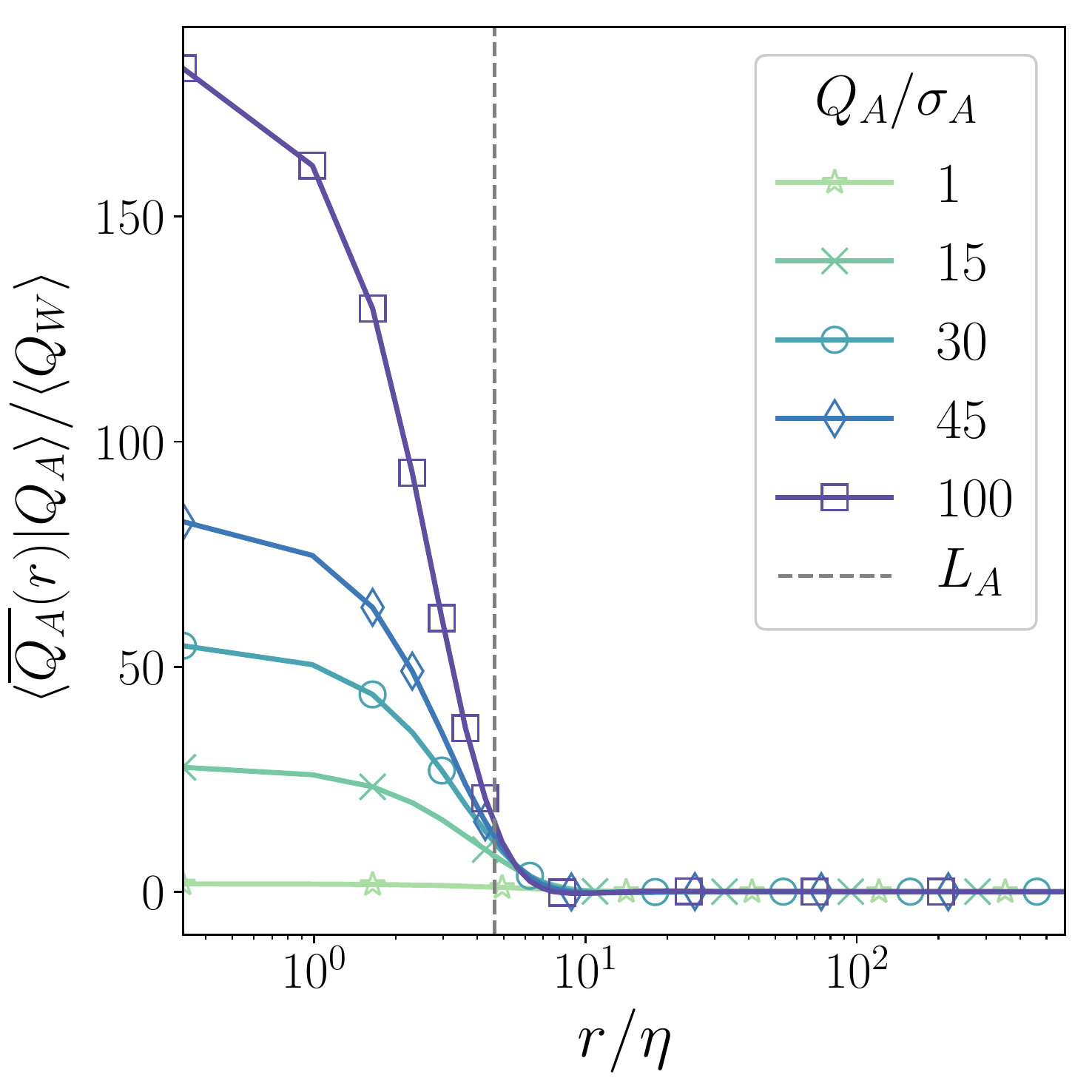}
 	 \caption{\label{fig: shell ave rad prof a} }
 \end{subfigure}%
  \begin{subfigure}{0.33\textwidth}
 	 \includegraphics[width=\textwidth]{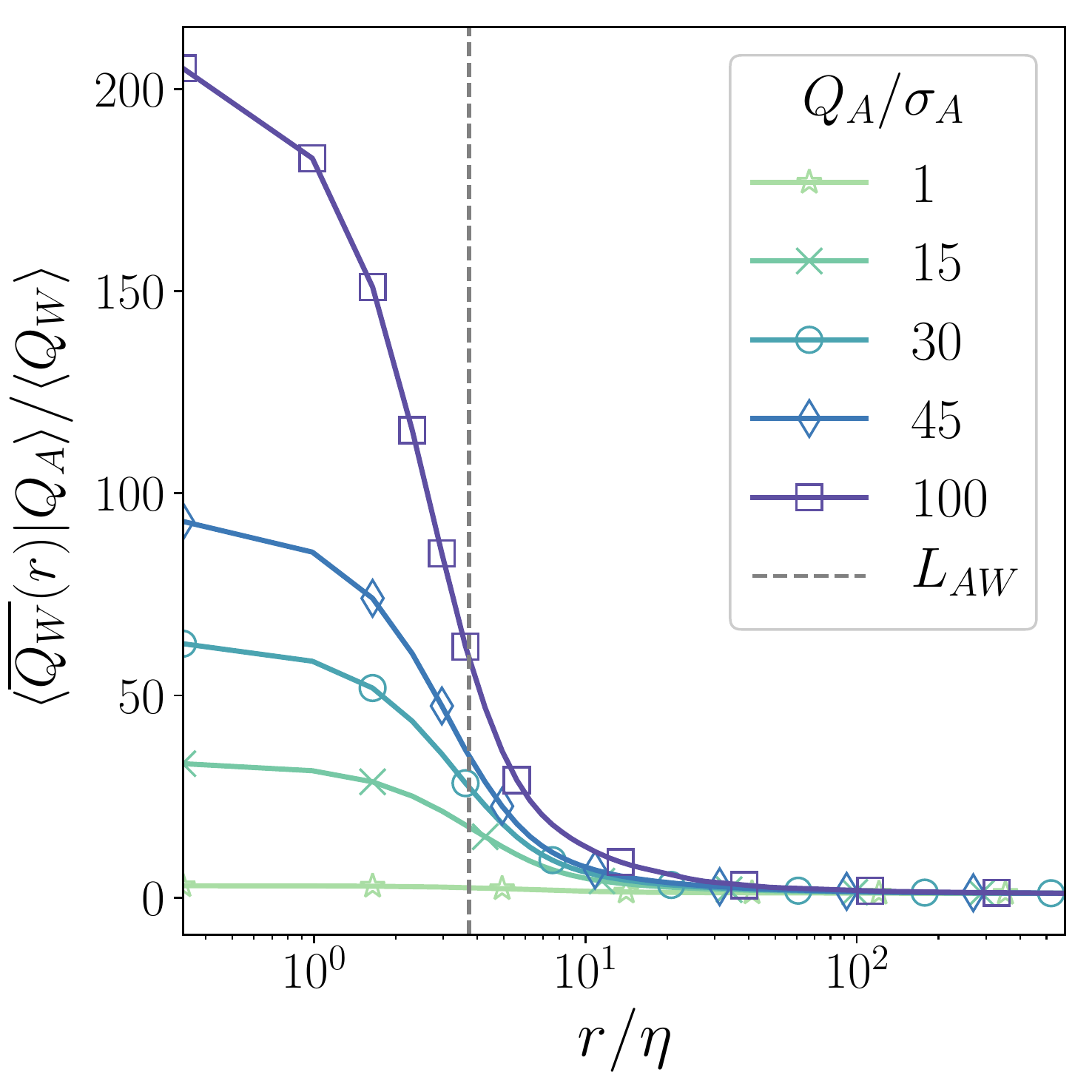}
 	 \caption{\label{fig: shell ave rad prof b} }
 \end{subfigure}%
  \begin{subfigure}{0.33\textwidth}
 	 \includegraphics[width=\textwidth]{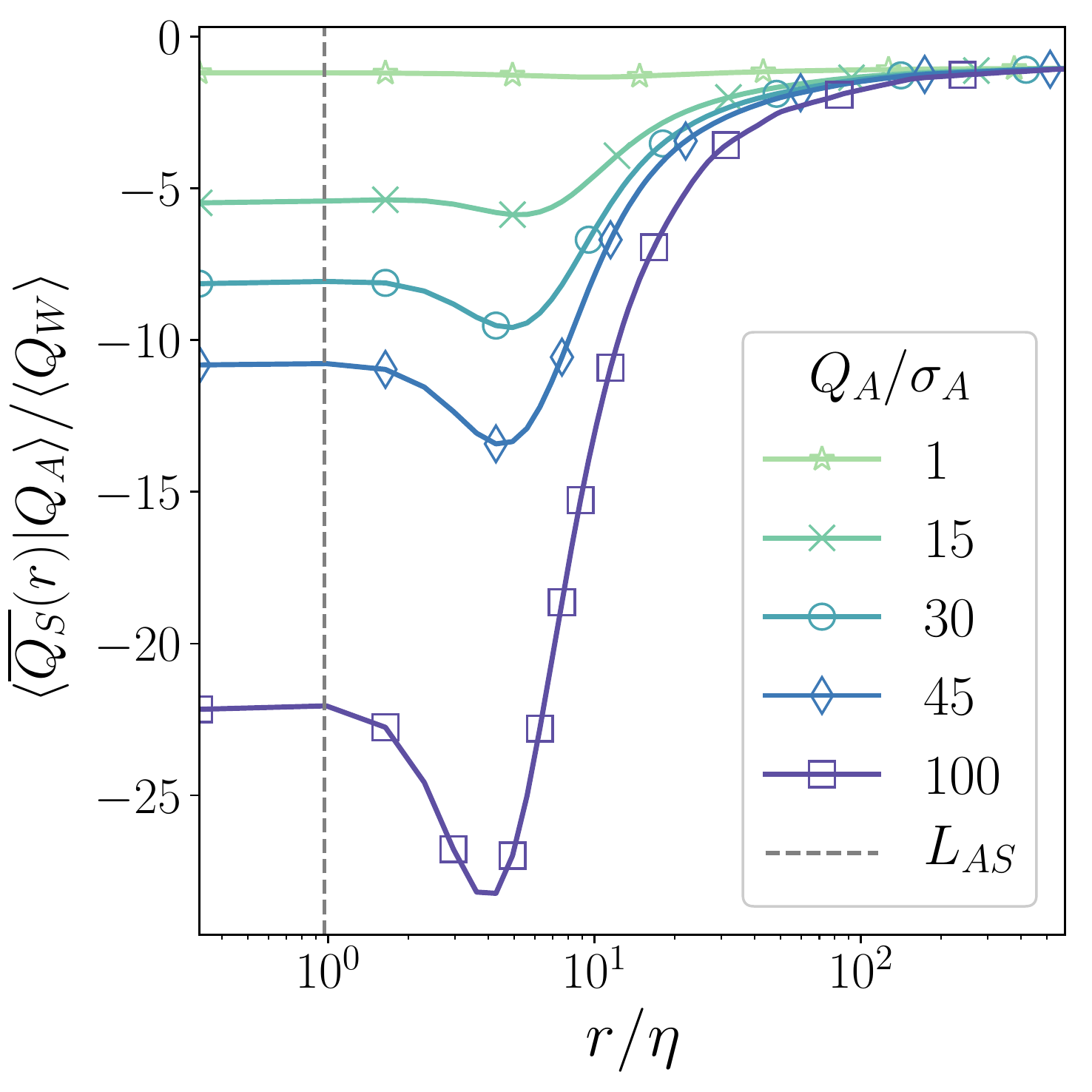}
 	 \caption{\label{fig: shell ave rad prof c} }
 \end{subfigure}%
 
  \begin{subfigure}{0.33\textwidth}
 	 \includegraphics[width=\textwidth]{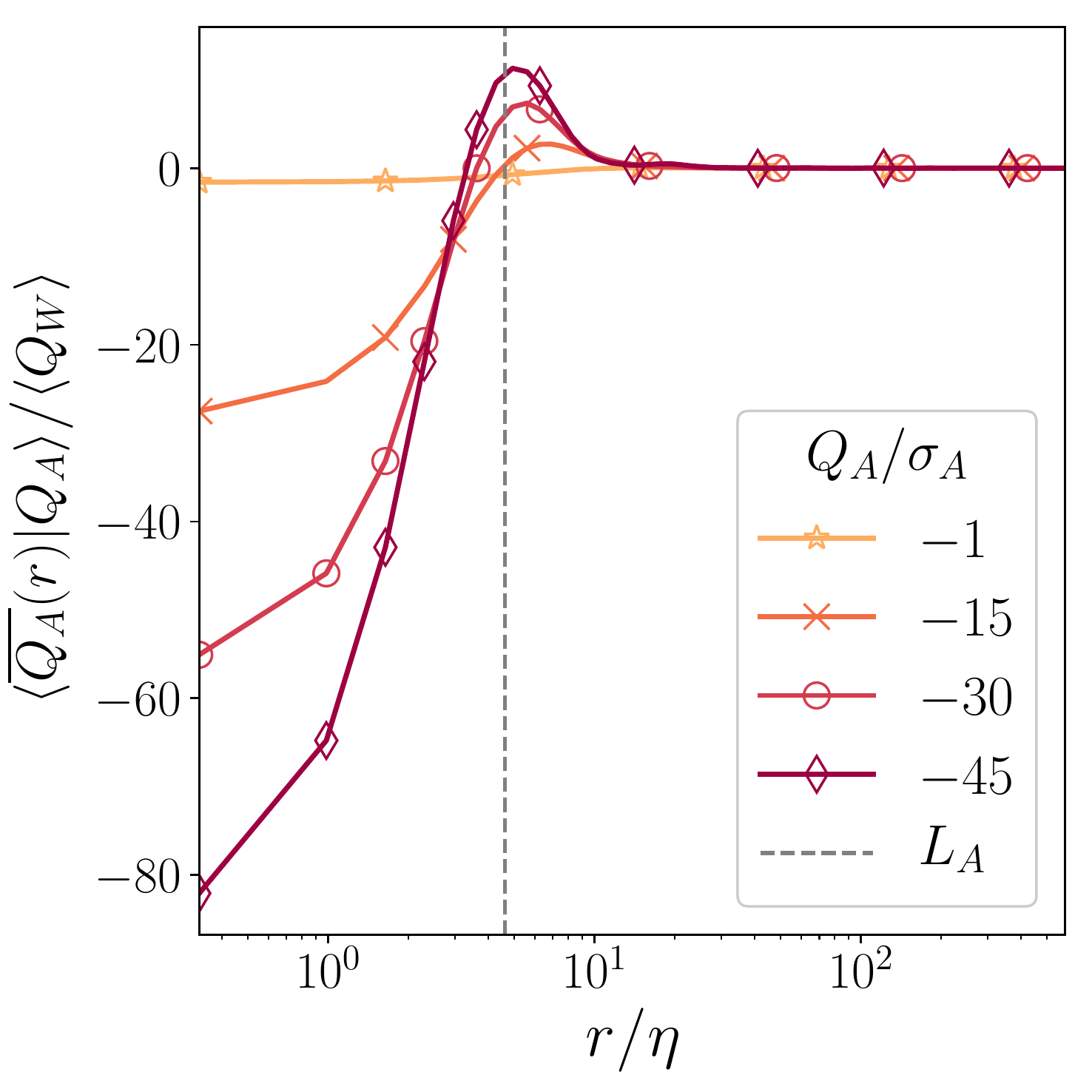}
 	 \caption{\label{fig: shell ave rad prof d} }
 \end{subfigure}%
  \begin{subfigure}{0.33\textwidth}
 	 \includegraphics[width=\textwidth]{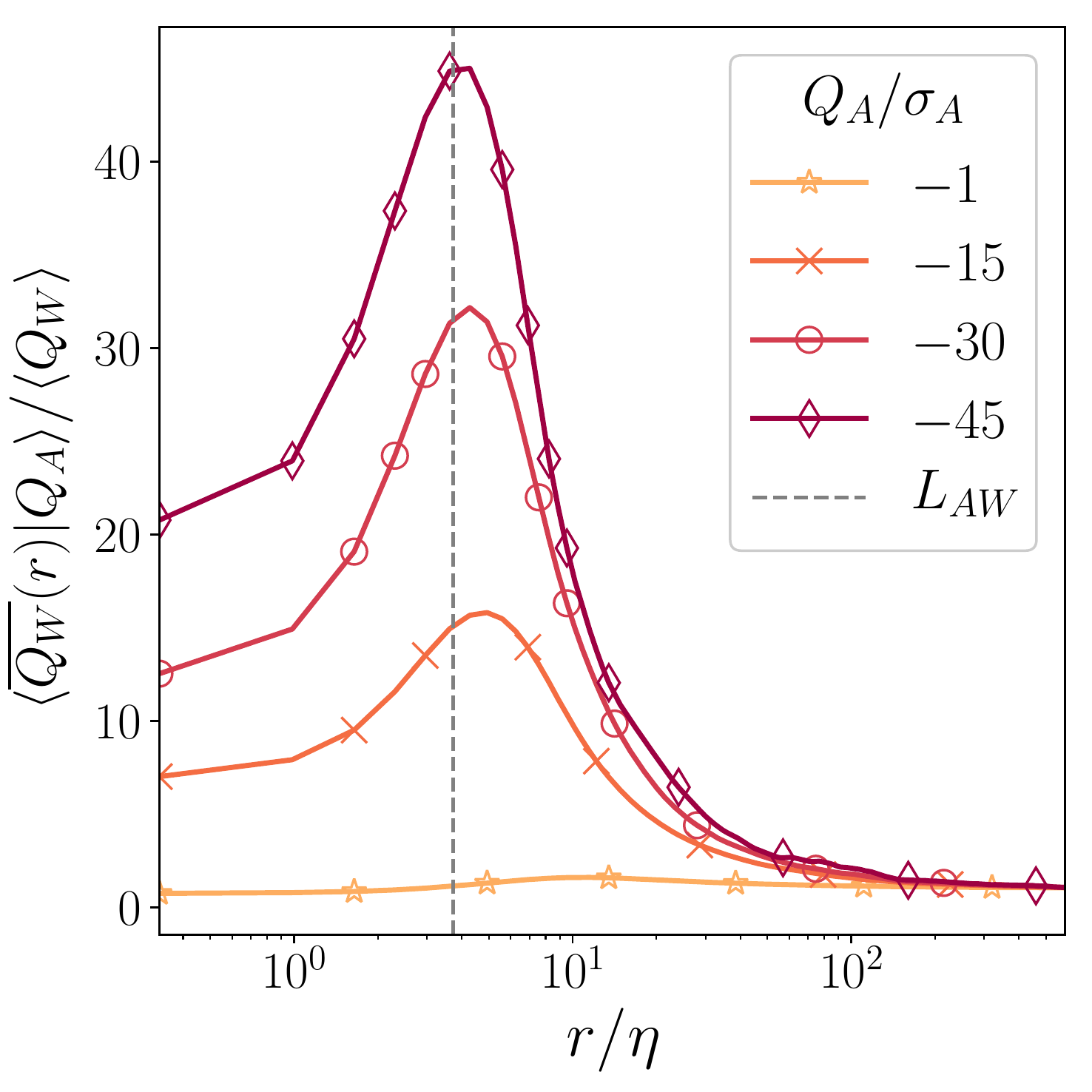}
 	 \caption{\label{fig: shell ave rad prof e} }
 \end{subfigure}%
  \begin{subfigure}{0.33\textwidth}
 	 \includegraphics[width=\textwidth]{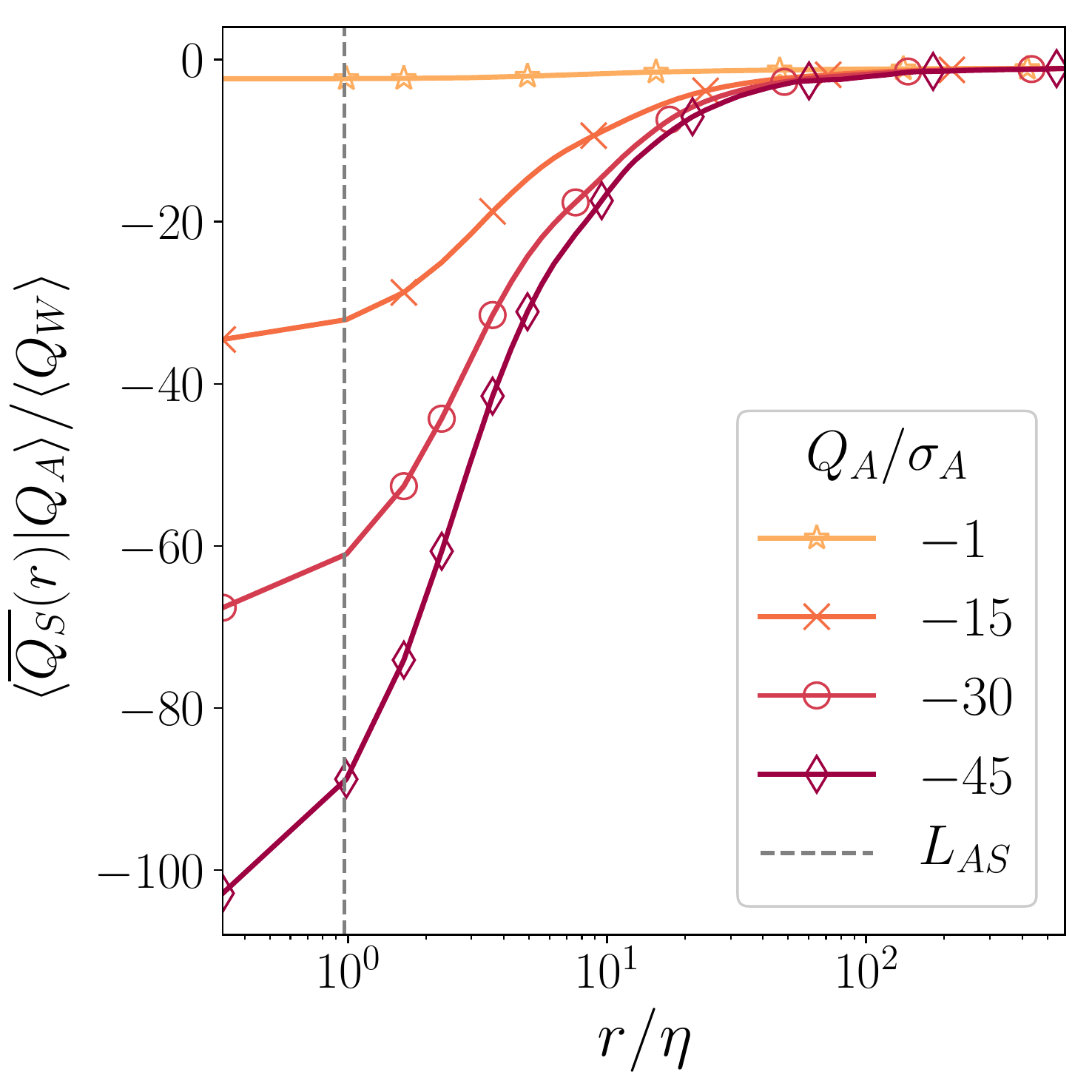}
 	 \caption{\label{fig: shell ave rad prof f} }
 \end{subfigure}%
	 \caption{
	    \label{fig: shell ave rad prof} 
	    \small
	    Radial profiles of $\overline{Q_A}(r)$ (\textit{a, d}), $\overline{Q_W}(r)$ (\textit{b, e}), and $\overline{Q_S}(r)$ (\textit{c, f}) at enstrophy-dominated regions $Q_A>0$ (\textit{a-c}) and strain-dominated regions $Q_A<0$ (\textit{d-f}).
	    The different colours show different intensity levels of the condition field, normalised to the respective standard deviation.
	    The vertical dashed lines show the correlation lengths associated with the auto- or cross-correlation functions derivable from the corresponding conditional averages.
	    The radial co-ordinate is normalised by the Kolmogorov length $\eta$ and the average contributions are non-dimensionalized by $\avg{Q_W}$, so as to be directly comparable across the different panels.
	    The data is taken from the simulation with $\Rey_\lambda = 168$.
	 }
\end{figure}

In particular, \cref{fig: shell ave rad prof} shows the profiles of $\cavg{\overline{Q_A}(r)}{Q_A}$,
$\cavg{\overline{Q_W}(r)}{Q_A}$ and $\cavg{\overline{Q_S}(r)}{Q_A}$ for a flow with  $Re_\lambda = 168$.
A selection of strain- and enstrophy-dominated conditions highlight the qualitatively different structure of these regions.
The figure shows a representative example of all analysed simulations.
All conditional averages are normalized by $\avg{Q_W}= -\avg{Q_S}$ as a measure of the background average.
To identify how extreme an event is, the conditions are given in terms of the standard deviation of $Q_A$.
Note that due to the skewness of the PDF of $Q_A$, the positive conditions extend to larger values. 
For reference, the PDFs of the velocity-gradient invariants are discussed in \cref{app: PDFs}; in particular the standard deviations of the fields are given in \cref{table: moments}.
In a way  \cref{fig: shell ave rad prof}  can also be considered as a model-free characterization of the isotropic component of the typical profiles of vortical and strain structures of given intensity.

\Cref{fig: shell ave rad prof a} illustrates the case of enstrophy-dominated reference points. It fully  supports the qualitative description of enstrophy-dominated regions suggested by the example of \cref{fig: turbulent sky}.
The initial value corresponds to the reference condition, i.e. the structures associated with the reference point.
For high-intensity conditions these are expected to be vortex tubes.
Note, however, that low-intensity conditions correspond to both quiescent regions and regions with a strong degree of cancellation between intense strain and enstrophy structures.
For all conditions the decay of the profiles indicates the length scale of the reference structures in the following sense.

Independent of the condition value, the decay is almost complete within approximately $10 \eta$ ( $\approx 2L_{A}$) for all conditions.  
So, all intense enstrophy-dominated profiles are contained within approximately the same length scale, i.e. they all decay completely before approximately the same radius.
A comparison with \cref{fig: shell ave rad prof b,fig: shell ave rad prof c}, shows that this is not due to the underlying vortical structures alone.
Both $\overline{Q_W}$ and $\overline{Q_S}$ conditional averages decay slowly for $r\gtrsim 2L_{A}$.
This can be associated with the gradual decorrelation of these fields with distance.
However, on these scales, for each condition value, $\overline{Q_W}$ must decorrelate at the same rate as $\overline{Q_S}$, so as to cancel out and leave virtually no imprint in the $\overline{Q_A}$ conditional averages in \cref{fig: shell ave rad prof a}.

The  $\overline{Q_W}$ and $\overline{Q_S}$ conditional averages in the slowly decaying outer regions are larger for more intense conditions. However, it is interesting to note that in relative terms (e.g. normalized to the peak of each conditional average), these regions are less prominent for more intense conditions. This reflects the notion that more intense structures are more localized than less intense ones.

Furthermore, there are regions of enhanced strain located at $r\approx L_{A}$ whose position appears to be fixed.
Their peak intensity is much lower than the one of the reference structure and  increases with the intensity of the condition. 
As a result the effective size and intensity of the enstrophy-dominated regions are attenuated. 
A similar purely kinematic result is described in \cref{app: analytic example} for an isolated vortex.
However, in the turbulent case the surrounding strain shells are relatively weak and much closer to the core of the central vortex region.
As a result they do not dominate their radial shells and the resulting enstrophy-dominated structures decay monotonically into the background.
(Note that this behaviour could also be accommodated by an isolated vortex with  axial instead of spherical symmetry.)

Conditioning on strain-dominated regions (cf. \cref{fig: shell ave rad prof d}) reveals a qualitatively different behaviour.
The conditional profiles change sign, unequivocally identifying strongly enstrophy-dominated regions around the reference strain-dominated structures.
Here the similarity to the isolated eddy example is even stronger, as the neighbouring vorticity structures actually dominate their radial shells.
The combined structures decay completely for $r\gtrsim 20\eta$. 
Again, this happens almost independently of the intensity of the condition, just as for the enstrophy-dominated profiles.
Considering the larger reference values of more intense structures, this leads to the conclusion that the more intense the $Q_A$ structures are, the more spatially localized they are.
Disentangling the individual strain and enstrophy contributions shows the reason for the different behaviour of $\cavg{\overline{Q_A}}{Q_A<0}$ (cf. \cref{fig: shell ave rad prof e,fig: shell ave rad prof f}).
The enstrophy-dominated shells also lie at $r\approx L_{A}$  for all large conditions, confirming this as the typical separation between intense strain and enstrophy structures.
However, their peak intensity is much closer to the peak intensity of the reference structures -- approximately half
(in contrast the peak of the strain-dominated regions in  \cref{fig: shell ave rad prof c} is approximately a  fifth of the reference condition).
Furthermore, the reference strain structures have much larger length scales and are much more diffuse than their enstrophy counterparts.

The correlation lengths associated with the corresponding correlation functions are also indicated in \cref{fig: shell ave rad prof}.
As noted, $L_A$ is a good indicator for the separation between regions of extreme vorticity and strain.
It also designates the length scale of $\overline{Q_A}$ around enstrophy-dominated points, on the one hand, 
and the radius of the enstrophy-dominated shell around strain-dominated points, on the other.
In all cases, this behaviour is robust for the investigated range of Reynolds numbers and forcing scales and types.
(Recall that $L_{A}$ scales with the Kolmogorov scale.)
In contrast, both $L_{A W}$ (shown in \cref{fig: shell ave rad prof b,fig: shell ave rad prof e}) and  $L_{A S}$ (shown in \cref{fig: shell ave rad prof c,fig: shell ave rad prof f}) grow much more rapidly with $\eta$.
In fact, with increasing $Re_\lambda$ their positions slowly drift to the right with respect to the peaks of the shown profiles. 
This behaviour is consistent with the Reynolds number dependence of the correlation lengths (cf. \cref{fig: length scale Re dependence}), which we can now explain.

It is easy to express the correlation lengths as averages of the conditional profiles, weighted by the corresponding PDFs and condition values. 
For example, 
\begin{align}
 L_{A S} =
 \frac{1}{\sigma_{S} \sigma_{A}}  \int_0^{\infty} \abs{ \int \cavg{\overline{Q_S}(r)}{Q_A=a} a \;\textrm{PDF}_{Q_A} (a) \;\mathrm{d}a }  \mathrm{d} r 
\end{align}
As the Reynolds number increases, the occurrence of extreme events increases which adds weight to the more pronounced high-intensity structures through the PDF.
The contributions from each conditional average also increases with $Re_\lambda$.
This is because, both the peak intensity and the length scales of the $\overline{Q_S}$ and $\overline{Q_W}$ profiles grow for each condition.
However, since the intensities and length scales of the strain and vorticity profiles increase at similar rates, the resulting  $\cavg{\overline{Q_A}(r)}{Q_A}$ profiles are practically unaffected. 
Hence the correlation length of $Q_A$ remains unaffected.

In summary, we find that the conditional averages of the velocity gradient invariants are dominated by kinematics not unlike that of an isolated vortex (\cref{app: analytic example}).
The central structure is associated with a surrounding shell of the opposite type, with separation of approximately $L_{A}$.
The intensity of the surrounding shell scales with the intensity of the reference structure.
The type of the central structure seems to play an important role in determining the shape and position of the surrounding shell.
Thus, strongly strain- and enstrophy-dominated regions consist of structures of similar characteristic scales of the order of $10\eta-20\eta$ independent of the reference intensity and $Re_\lambda$.
They originate from underlying strain and enstrophy structures with much larger and more widely varying length scales. 
However, the outskirts of these structures always share similar decay profiles and hence cancel out.

\section{Degree of pressure (non-)locality}
\label{sec: pressure}
\subsection{Conditional pressure contributions}
In order to translate the structure of the velocity gradient to the pressure field, we need to
examine the effect of the Poisson kernel.  So,
 we proceed by considering the conditional averages of the contributions to the pressure field $\avg{\overline{q_A}(r)| Q_A}$.
Using \cref{eq: rad p check}, it is immediate that
\begin{align}
\avg{p(\vx)|Q_A(\vx)}  \equiv \cavg{p}{Q_A} =  \int_0^{\infty} \avg{\overline{q_A}(r)| Q_A}\mathrm{d} r.
\label{eq: p cond check}
\end{align}
Recall from \cref{eq: qA def} that after angle averaging the kernel weight is proportional to $r$.
Clearly then the Poisson kernel separates the parts of $\cavg{\overline{Q_A}(r)}{Q_A}$ which decay faster  from the ones which decay more slowly than linearly.
Thus, the local extrema of $\cavg{\overline{q_A}(r)}{Q_A}$ occur at the isolated points where  $\cavg{\overline{Q_A}(r)}{Q_A}$ decays as $1/r$.
Additionally, the Poisson kernel has a more pronounced effect on more intense conditions, 
because they are associated with profiles with larger amplitudes.
Moving beyond these general consideration for turbulent flows is analytically challenging.
To aid the intuition the simple case of the contributions to the pressure at the centre of an isolated Townsend eddy is described in \cref{app: analytic example}.

\begin{figure}
 \begin{subfigure}{0.5\textwidth}
 	 \includegraphics[width=\textwidth]{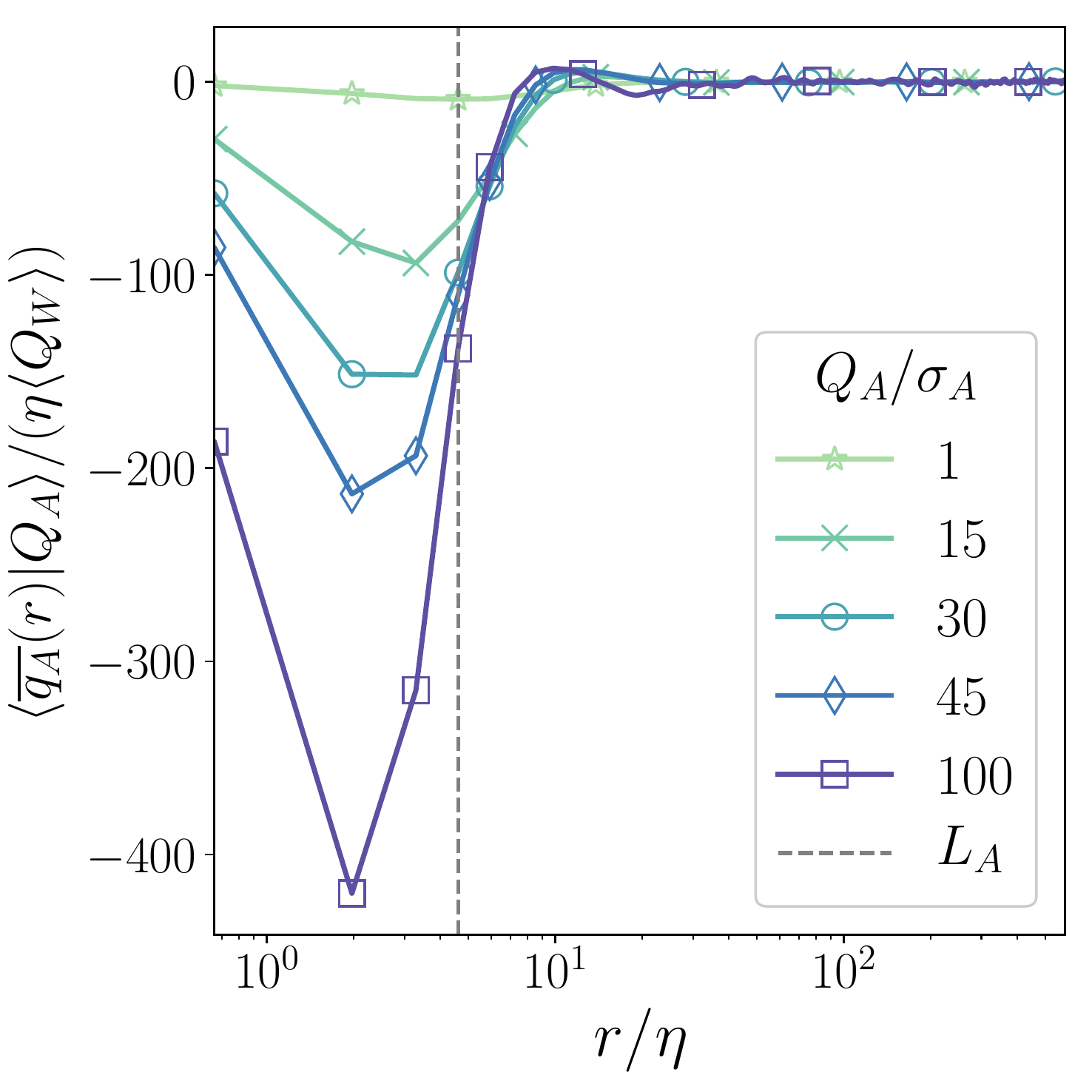}
 	 \caption{\label{fig: p_contrib_rad_prof a} }
 \end{subfigure}
 \begin{subfigure}{0.5\textwidth}
 	 \includegraphics[width=\textwidth]{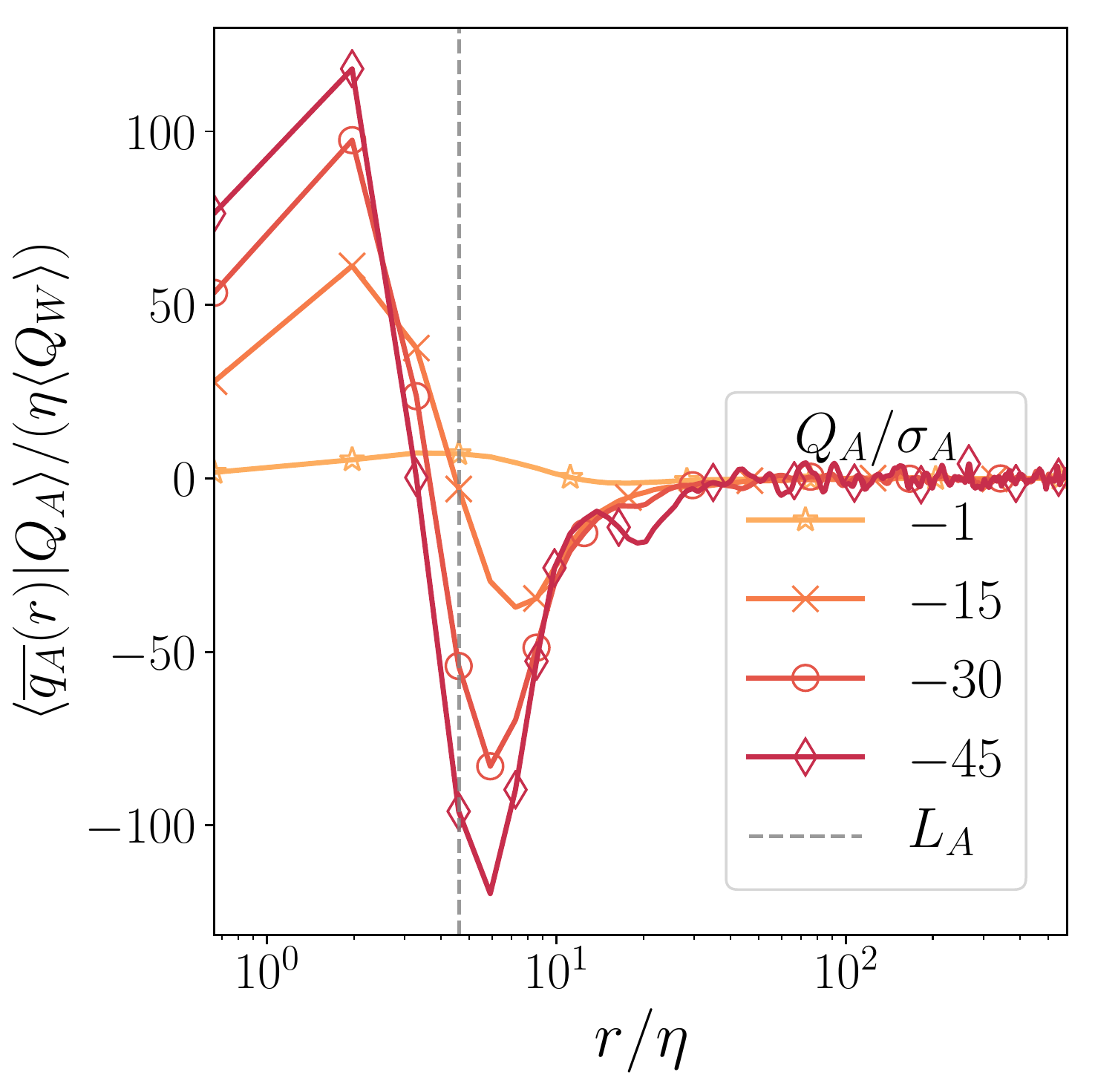}
 	 \caption{\label{fig: p_contrib_rad_prof b} }
 \end{subfigure}
	 \caption{
	    \label{fig: p_contrib_rad_prof} 
	    \small Average contributions to the pressure at (\textit{a})  enstrophy-dominated regions  and (\textit{b})  strain-dominated regions from thin spherical shells of prescribed radius.
	    The different colours show different intensity levels of the condition field, normalised to the respective standard deviation.
	    The radial co-ordinate is normalised by the Kolmogorov length $\eta$ and the average contributions are non-dimensionalized by $\eta \avg{Q_W}$.
	    The data is taken from the simulation with $\Rey_\lambda = 168$. 
	 }
\end{figure}

To address the turbulent case we consider a representative example of $\cavg{\overline{q_A}(r)}{Q_A}$ for a range of conditions for $Re_\lambda = 168 $ (see \cref{fig: p_contrib_rad_prof}).
Based on the effect of the Poisson kernel, three qualitatively distinct regions can be identified for all condition types and intensities: a core, a tail and an intermediate region.
The central core for all conditions is dominated by the Poisson kernel. 
The relatively constant $\cavg{\overline{Q_A}(r)}{Q_A}$ central profiles in this region lead to growing pressure contributions, which peak at the edge of the core.
Naturally, the height of the peak increases with the intensity of the reference structure.
In the analysed simulation, it occurs at ($r \approx 2 \eta$).
So, in view of their impact on the pressure field, the cores of both enstrophy- and strain-dominated structures may be defined as the regions where the $\overline{Q_A}$ conditional averages decay more slowly than $1/r$. 
It should be pointed out that an increase in small-scale resolution may lead to steeper gradients at these scales and consequently reduce the size of these regions.
Comparing simulations 6 and 7 reveals this to be the case for the strain-dominated, but not for the enstrophy-dominated conditions.

In contrast, the tails of the profiles are dominated by the far-field asymptote of $\overline{Q_A}$.  
Here the type of condition begins to play a role. 
As shown in the figure, enstrophy-dominated conditions receive almost no net pressure contributions
from regions beyond $ \approx 10 \eta$.
In comparison, for strain-dominated conditions this happens at $\approx 20 \eta$.
In both cases however, on these scales the Poisson kernel does little more than to
slow down the decay rate of the high intensity $\overline{Q_A}$ conditional averages
and  enhance the unavoidable statistical noise.
It should be noted that this result is entirely contingent on the fact that $\avg{Q_A}=0$.
If the mean were finite, the resulting pressure contributions would grow linearly with radius and thus in a physical system would be dominated by the largest scales and/or boundary conditions(unless those also cancel each other).
This is in fact the case for the individual strain and enstrophy contributions to the pressure, quantified by the conditional averages of $-2 r \cavg{\overline{Q_S}(r)}{Q_A}$ and $-2 r\cavg{\overline{Q_W}(r)}{Q_A}$ (not shown).
It is because of the global balance of $Q_S$ and $Q_W$ ($\avg{Q_S}=-\avg{Q_W}$) that the dominant pressure contributions at extreme events remain relatively local.

The intermediate region in between the core and the tail of the conditional averages is where the properties of the velocity gradient and the Poisson kernel interact most interestingly and yield the bulk of the pressure contributions.
In this region, for all types of conditions, larger intensity corresponds to steeper profiles.
This suggests that the pressure at more intense structures is more local, independent of the type of the structure.
More striking, however, are the differences between the enstrophy- and strain-dominated reference structures.

On the one hand, the enstrophy-dominated structures simply decay into the background (faster than $1/r$),
so the Poisson kernel simply slows down that decay.
It also reveals a small footprint of the surrounding strain regions at $r\approx 10\eta$ for the very extreme conditions.
However its amplitude is negligible for the examined range of Reynolds numbers.
Overall, this leads to the conclusion that the pressure at strongly enstrophy-dominated structures is a very local quantity, with most contributions coming from a shell of radius $\leq L_{A}$.
Furthermore, due to the Poisson kernel, the periphery and the core of the structures have comparable weights.
Recall that the structure and spatial scale of the periphery is sensitively determined  by the cancellation of strain and enstrophy contributions.
In that sense, strongly enstrophy-dominated regions exhibit a form of self-shielding with a characteristic length scale of $L_{A}$.

On the other hand, the pressure contributions of the  strain-dominated conditional averages are determined primarily by the presence of the weakly enstrophy-dominated shell at $r\approx L_{A}$ (cf. \cref{fig: shell ave rad prof d}). 
The Poisson kernel amplifies the importance of this region, so much so that it overwhelms the pressure contributions from the strain-dominated core (cf. \cref{fig: p_contrib_rad_prof b}).
This leads to the negative pressure values at intense strain regions noted by e.g. \cite{Yeung2012}.
As a result, the central core has a direct impact only on the magnitude of the corresponding pressure.
The length scale of the pressure contributions, however, is governed by the structure of the surrounding enstrophy-dominated regions.
So, in strain-dominated regions the self-shielding mechanism is unbalanced and the pressure is determined at
larger length scales, which reach the inertial range. In all analysed simulations the inertial range is considered to extend to several tens of $\eta$.

\subsection{Length scale of non-locality}

In order to set an upper bound on the non-locality of the pressure contributions more  precisely, we introduce a concrete length scale, $r_\alpha(Q_A)$.
We refer to it as a threshold radius. 
Explicitly it is defined as 
\begin{align}
  \label{eq: rad def}
    r_\alpha(Q_A) = \underset{d}{\max} \left\{ \left\lvert \int_{0}^{d}  \cavg{\overline{q_A}(r)}{Q_A} \mathrm{d} r   \right\rvert = \alpha \left\lvert \cavg{p}{Q_A} \right\rvert \right\}.
\end{align}

In essence, for each condition $Q_A$, we compute the size of the local neighbourhood which contributes a fraction $\alpha$ of the conditional pressure average at the reference point. 
If cancellations lead to the existence of several such (nested) neighbourhoods, we take the largest one.

\begin{figure}
  \centering
 \begin{subfigure}{0.52\textwidth}
	 \includegraphics[width=\textwidth]{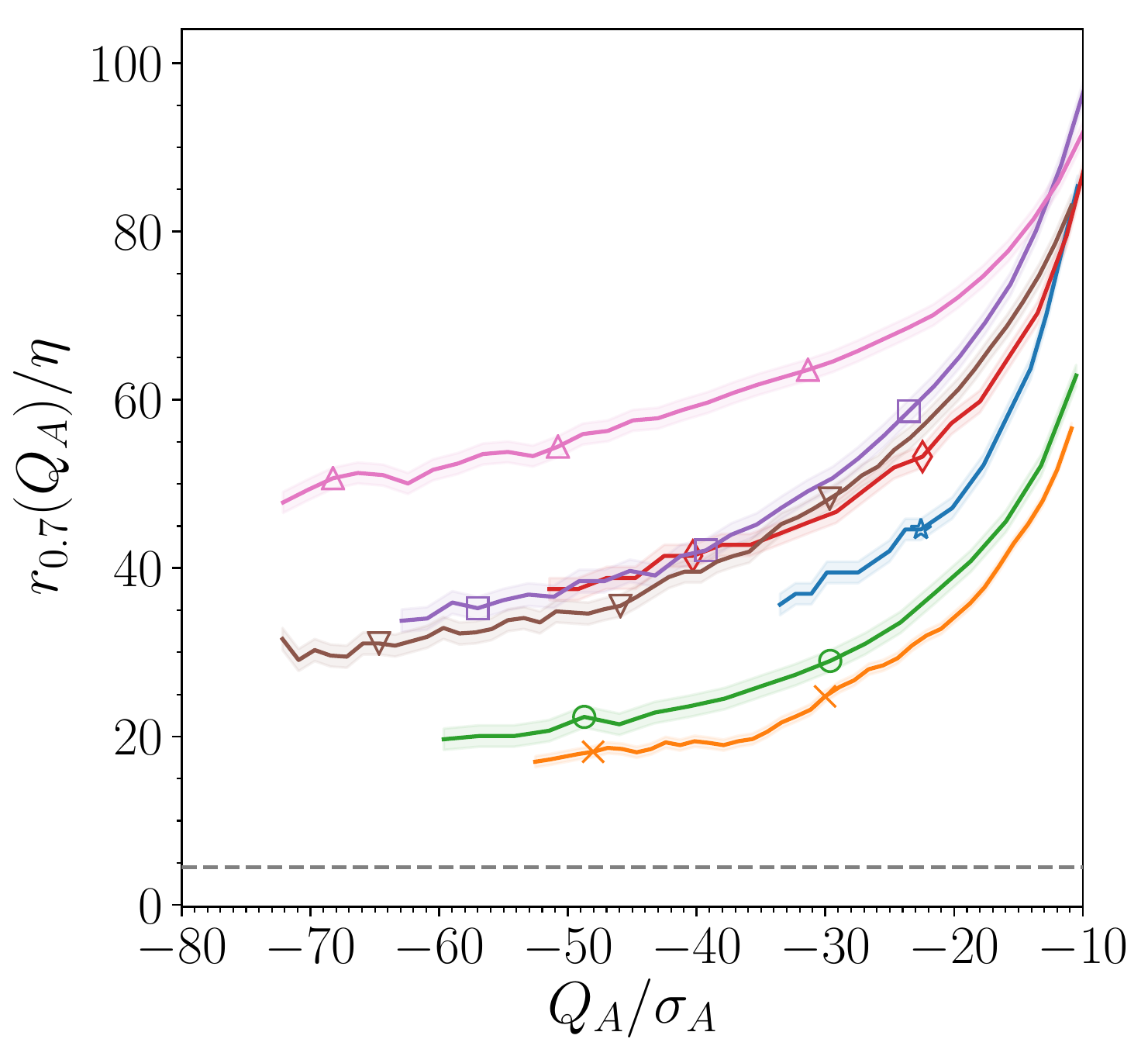}
	 \caption{\label{fig: thresh rad a} }
 \end{subfigure}%
 \begin{subfigure}{0.5\textwidth}
	 \includegraphics[width=\textwidth]{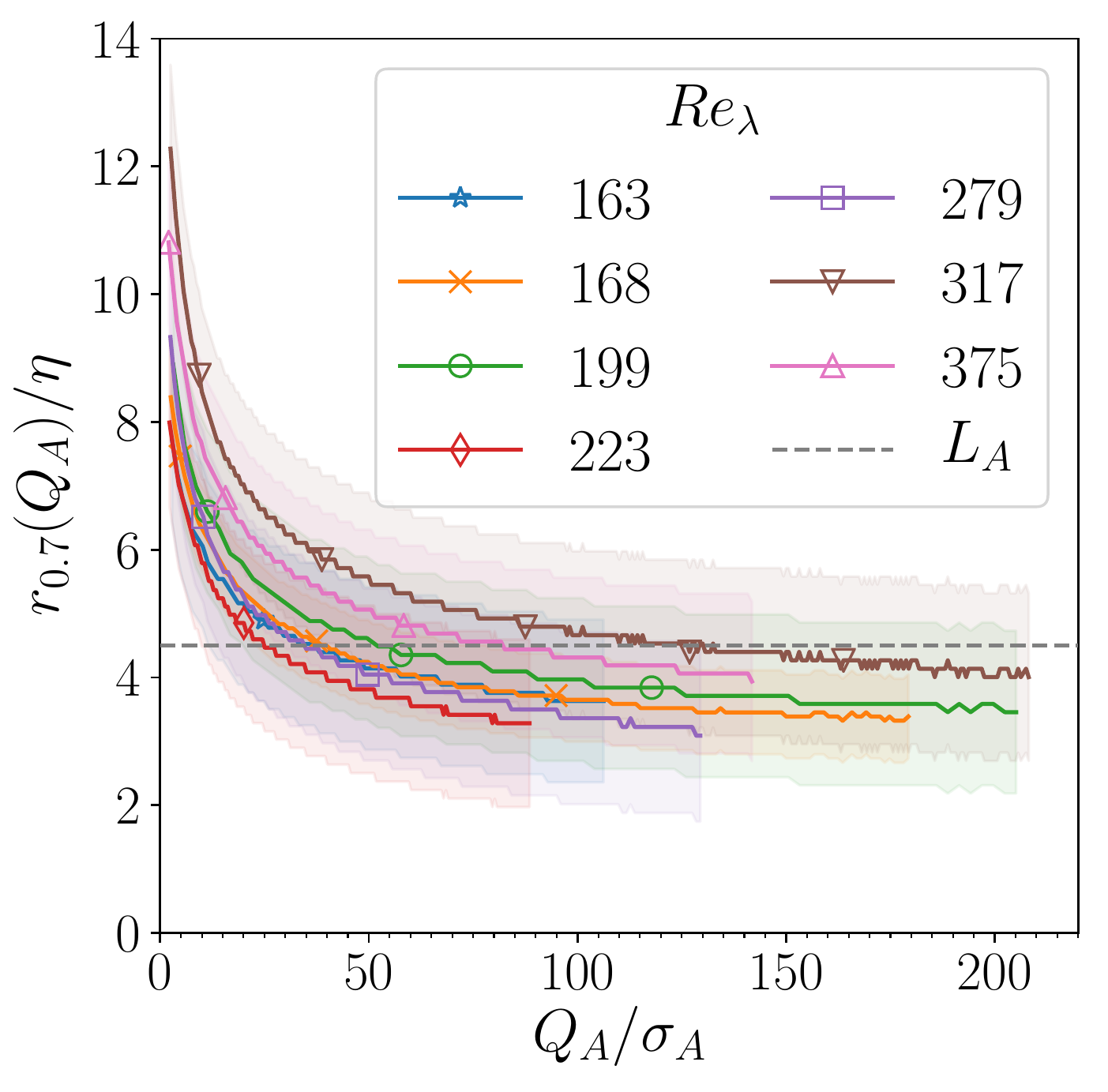}
	 \caption{\label{fig: thresh rad b} }
 \end{subfigure}%
	 \caption{
	    \label{fig: thresh rad} 
	    \small Threshold radius $r_{0.7}(Q_A)$ for a range of Reynolds numbers, for strain-dominated conditions (\textit{a}) and enstrophy-dominated conditions (\textit{b}).
	    The shaded regions designate an estimate for the numerical and statistical uncertainty of the computation due to the cell width.
	    The dashed line at $4.5\eta$ corresponds to the best fit value for the integral correlation length $L_{A}$.
	    The low-intensity conditions $-10\sigma_A<Q_A<\sigma_A$ are omitted, since for those the threshold radius is dominated by statistical and numerical uncertainties.
	 }
	\end{figure}

In the following, we consider specifically a threshold level of $70\%$, i.e. $r_{0.7}(Q_A)$.
This threshold is chosen to reduce the contamination from the statistical and numerical noise for all datasets and all conditions fields.
The results for all examined simulations are plotted in \cref{fig: thresh rad}. 
The shaded regions represent an estimate of the numerical uncertainty induced by the finite cell width.
In order to clarify the dependency on the condition, the plotted curves are obtained after oversampling the $\cavg{\overline{q_A}(r)}{Q_A}$ profiles by a factor of ten with respect to $r$.

As anticipated, for any given Reynolds number, the threshold radius decreases as the intensity of the condition value increases. 
This supports and quantifies the conclusion drawn above that the pressure becomes increasingly more local in regions of extreme velocity gradients.
The strong  dependence on the type of conditioning is also evident.

For enstrophy-dominated conditions, see \cref{fig: thresh rad b}, $r_{0.7}(Q_A)$ quickly 
decays to a few Kolmogorov scales at small condition intensities and could be crudely approximated by the correlation length  $L_{A}$ for the more intense conditions. 
Thus, the conditional pressure average at these conditions is determined by a dissipation-scale region of the size of the reference enstrophy-dominated structure.
From this it follows naturally that $r_{0.7}(Q_A)/\eta$ should not depend strongly on the Reynolds number.
While the data give some hints to a small Reynolds number dependence, the issue can be settled by improved small-scale resolution simulations. 
These would also be necessary to uncover any significant dependence on the intensity of the condition for the most extreme vortices.

For strain-dominated conditions, see \cref{fig: thresh rad a}, the minimum value of the threshold radius is in the tens of $\eta$.
This quantifies the importance of the surrounding enstrophy-dominated shell described above.
Clearly, the Poisson kernel plays the crucial role of extending the range of significant pressure contributions to inertial scales.
As a result, they can no longer be tied to a single velocity-gradient structure and its kinematics, but are determined by the spatial distribution of a multitude of such structures.
Thus the threshold radius for these conditions cannot be expected to  naively scale with $\eta$.
There are indications in the data that it increases with Reynolds number.
However, several simulations (including the ones with different forcing scale and type) break the trend, suggesting a possibility for non-universal behaviour. 
Critically, the significant difference between the simulations with the similar $Re_\lambda$ but different small-scale resolution indicates that, for the strain-dominated conditions, the threshold radius should be considered only as an upper bound, which may decrease with improved small-scale resolution.

For completeness, we should mention that we have omitted from consideration the conditions $-10\sigma_A <Q_A< \sigma_A$.
For them the threshold radius is dominated by statistical and numerical uncertainties, which is not entirely unexpected.
On the one hand, at the strain-dominated end of this range, there are nearly vanishing local pressure contributions because of almost perfect cancellation between the contributions from the strain-dominated core and the surrounding enstrophy-dominated shell. This dramatically increases both statistical and numerical uncertainties in the computation of the threshold radius.
On the other hand, the low-intensity conditions correspond to the large quiescent regions with ambient or slightly positive pressure (as seen in \cref{fig: visualisations}).
 Because these quiescent regions occupy a significant volume of the simulation domain (see \cref{app: PDFs}),
it follows from homogeneity  that for them  $\avg{p|Q_A} \approx \avg{p}=0$.
They are also not associated with well-defined structures.
So, for them the pressure contributions vanish at all radii, for instance see the $Q_A= \pm \sigma_A$ profiles in \cref{fig: p_contrib_rad_prof}.
As a result, analytically the threshold radius is expected to diverge in these regions.

\section {Discussion and conclusions}
To summarize, we presented a numerical study characterizing the small-scale structure of homogeneous and isotropic, statistically stationary turbulence in the range $\Rey_\lambda \in [160,380]$.
We computed characteristic length scales associated with the second invariant of the velocity gradient, $Q_A$, its strain and enstrophy components and their separation.
This led naturally, through \cref{eq: Greens soln}, to an effective limit on the non-locality of conditional pressure statistics.

As a first diagnostic of the velocity gradient, we computed the two-point correlations of $Q_A$, $Q_S$ and $Q_W$.
The characteristic length scale of $Q_A$ structures, as measured by the corresponding correlation length, is of the order of $4.5\eta$ and is approximately independent of Reynolds number.
This can be understood by considering its strain and enstrophy components.
Firstly, enstrophy and strain structures are strongly entangled (cf. figures \labelcref{fig: visualisations}(\textit{a}-\textit{c})), and exhibit identical correlations on inertial and integral scales (cf. \cref{fig: SW corr,fig: length scale Re dependence}). 
This leads to cancellations and reduces the scale of $Q_A$.
Secondly, while their correlation lengths (of the order of tens of the Kolmogorov scale) grow with $Re_\lambda$, 
they do so at such rates that the resulting $Q_A$ structures retain their characteristic length scale. 
This behaviour is supported by the kinematic relationship between the rate-of-strain and the vorticity \citep{Ohkitani1994}.
Simply put, the $Q_A$ correlations are the small-scale residue, which remains from cancellation of enstrophy and strain structures on inertial and integral scales.

To characterize in more detail the different types of turbulent structures, we considered the angle-averaged two-point conditional mean profiles of $Q_A$ around strain- and enstrophy-dominated reference points of given intensity.
This way we could describe simply the isotropic component of prototypical vortex tubes and strain sheets of different intensities and the typical background in which they are embedded.
The analysis implies that regions strongly dominated by either enstrophy or strain exhibit quickly decaying profiles.
Thus, the theoretically expected homogenization of $\avg{Q_A}$ (cf.  \cref{eq: omega limit}) is achieved at scales of only a few $L_{A}$, independent of the type of reference point.
This already places an upper bound on the scale of pressure non-locality for these regions, which are associated with the negative tail of the pressure PDF.

More locally, intense enstrophy-dominated regions are organized in small-scale vortex structures which generate surrounding envelopes of comparatively weaker strain. 
Equivalently, intense strain-dominated regions are typically found in the neighbourhood of comparatively stronger vortical structures. 
The typical separation between the two is approximately $L_{A}$,  as implied by the position of the peaks in the $\cavg{\overline{Q_W}(r)}{Q_A<0}$ and $\cavg{\overline{Q_S}(r)}{Q_A>0}$ profiles (cf. \cref{fig: shell ave rad prof c,fig: shell ave rad prof e}).
Moreover, it is almost independent of the intensity of the condition for intense conditions.
The enstrophy-dominated $Q_A$ structures are typically a weakly attenuated and more compact version of the underlying enstrophy. They are embedded in a flat background due to the relative amplitude and separation of the underlying enstrophy and strain structures.
In contrast, the strain-dominated structures  retain a clear signature of the associated neighbouring vortex (cf. \cref{fig: shell ave rad prof d}).
Finally, the correlation length $L_{A}$ characterizes well both enstrophy- and strain-dominated conditional profiles, albeit in different ways.
For the former, it gives a proxy for the length scale of decay, while for the latter it indicates the position of the surrounding enstrophy-dominated structure.

Motivated by the numerous analogies between the Poisson equation describing the pressure field and the one for the electric potential in electrostatics or plasmas,
the second aim of this study is to describe and quantify on the non-locality of the pressure.
In particular, we considered the locality of the contributions to the single-point conditional pressure average.
Since the pressure field is sourced by $Q_A$, only the effect of the Poisson kernel remains to be applied to the above description of the $Q_A$ structures (cf. \cref{eq: p cond check}).

The analysed simulations imply that (cf. \cref{fig: p_contrib_rad_prof}), for all condition types and intensities the large-scale contributions to the pressure are small in absolute terms.
Note that this is unrelated to the existence of large-scale pressure--pressure correlations.
It is simply due to the homogeneity of the flow, and specifically of $Q_A$, which nullifies the effect of the Poisson kernel at large scales.
However, the \emph{relative} importance of the different scales depends strongly on the intensity of the local neighbourhood.
In general, we find that the pressure at regions with more intense velocity gradients is determined more locally and \textit{vice versa}.
Furthermore, the type of reference point qualitatively changes the locality of the pressure,
because the Poisson kernel strongly amplifies the difference in the $\overline{Q_A}$ profiles around enstrophy- and strain-dominated points.
As a result, the pressure  in enstrophy-dominated regions is determined primarily by the local structure 
and is thus dominated by a Kolmogorov-scale neighbourhood. 
In contrast, the pressure in strain-dominated regions is determined by a larger neighbourhood, which may extend to inertial scales, 
because the pressure contributions of the local strain structure are cancelled out and overtaken by the effect of nearby vortices.
From another perspective, this suggests that strong vortices dominate or at least strongly influence the pressure in their neighbourhoods up to inertial scales
(of the order of tens of the Kolmogorov scale).

An upper bound on the non-locality of the pressure contributions is provided by the threshold radius $r_{0.7}(Q_A)$, shown in \cref{fig: thresh rad}.
For enstrophy-dominated conditions, it shows approximate independence of $Re_\lambda$, given the small-scale resolution of the data.
For most strain-dominated conditions, a tentative trend of increasing non-locality with $Re_\lambda$ is observed in the simulations with $k_\textrm{max} \eta \approx 2$
despite difference in the large-scale forcing.
However, a comparison between simulations with different small-scale resolution and similar Reynolds numbers
suggests that numerical effects strongly increase the threshold radius.
This has little impact for the enstrophy-dominated regions because of the large core contributions. 
However, the cancellations associated with the core of strain-dominated regions render the calculations there much more sensitive.
Thus, further simulations with improved small-scale resolution and larger statistics might be necessary for more quantitative statements about those.

From these observations emerges a description for turbulent shielding in partial analogy to the one in plasmas.
A strongly enstrophy-dominated region acts like a strong negative charge embedded in a weaker, positively charged cloud, 
i.e. a strain-dominated region.
The resulting pressure field is analogous to the electric potential in plasmas.
Unlike in plasmas however, the charge carriers are neither uniformly distributed, nor isotropic.
So, the positive cloud is not strong enough to nullify the effect of the negative charge but merely attenuates it.
It is plausible that this is due, at least in part, to the different anisotropic structure and length scales of the vortex tubes and strain sheets.
In any case,  enstrophy-dominated regions determine their pressure locally and also influence strongly the pressure at the surrounding strain-dominated regions.
So, regions of intense velocity gradients, i.e. intense turbulence, correspond to regions with a lot of elementary charges of both polarities. They tend to have an overall negative potential, i.e. pressure.
To maintain overall balance this is cancelled out by a smooth diffuse positive component which corresponds to the large-scale featureless quiescent turbulent voids.
In the quiescent regions the homogeneity of the flow dictates that the pressure is positive and determined non-locally by integral-scale contributions.
(Or in the analogy, the overall charge neutrality is responsible for that.)
In this analogy the correlation length of $Q_A$  would be akin to the Debye length for regions of intense turbulence.

A still open question which bears deeper investigation is the origin of the above-described shielding phenomenon. 
We noted that, in the considered flow the shielding at very large scales is purely a result of statistical homogeneity.
However, on inertial and dissipation scales there are several effects which can contribute to the shielding.
Firstly, there is the purely kinematic generation of strain from enstrophy (and \textit{vice versa}), as applied to individual structures. 
Secondly, there is the spatial distribution of the extreme structures themselves, which is encoded by the dynamics of the Navier-Stokes equations. 
Thirdly, there is the different anisotropic structure of vortex tubes and strain sheets.
The presented analysis cannot address the last point.
However, it does seem to suggest that the kinematic effect is more relevant to the pressure around strongly enstrophy-dominated regions, while the dynamic effect plays a larger role for the pressure at strain-dominated regions. 
Further analysis in the spirit of, for instance, the local versus non-local strain decomposition proposed by \cite{Hamlington2008} may shed further light on this question.

\acknowledgements{
We gratefully acknowledge insightful discussions with Dr~J.~Lawson and Dr C.~C.~Lalescu, as well as the generous assistance of
Dr~C.~C.~Lalescu in the computational implementation of the analysis. 
We would like to thank the anonymous reviewers for the insightful and constructive criticism.
The results were analysed in part with the computational resources of the Max Planck Computing and Data Facility.
The authors gratefully acknowledge the Gauss Centre for Supercomputing e.V. (www.gauss-centre.eu) for funding this project by providing computing time on the GCS Supercomputer SuperMUC at Leibniz Supercomputing Centre (www.lrz.de).
We are grateful for the financial support of the Max Planck Society.
}

\appendix

\section{Analytic example of a model eddy}

\label{app: analytic example}
As a first step to understanding the spatial structure of the $Q_A$ field  and its impact on the pressure non-locality consider a toy example for the structure at the reference point --- a Townsend eddy,  a localized blob of concentrated vorticity
\citep[p.~370]{Davidson_book}.
The vorticity is described in cylindrical polar coordinates $\vx(\rho,\theta, z)$ by 
\begin{align}
\v{\omega}(\vx) = u_\theta(\vx) \bra{ \frac{z}{r_c^2} \v{e}_r + \bra{\frac{2}{\rho}- \frac{\rho}{r_c^2}} \v{e}_z},
\end{align}
with angular velocity $u_\theta (\vx) = \Gamma   \rho \exp \bra{-\vx^2/ 2 r_c^2 }$ and characteristic length scale  $r_c$. 
The associated quadratic invariants are given by
 \begin{align}
 \label{eq: Townsend eddy QS}
 Q_S(\vx)   &= -\Gamma^2 \frac{\vx^2 \rho^2 }{4 r_c^4} \exp \bra{- \vx^2/r_c^2}\\
 \label{eq: Townsend eddy QW}
 Q_W(\vx) &= \Gamma^2\bra{\frac{\vx^2 \rho^2 }{4 r_c^4}- \frac{\rho^2}{r_c^2}+ 1 } \exp \bra{- \vx^2/r_c^2}\\
 \label{eq: Townsend eddy QA}
 Q_A(\vx) &=\Gamma^2 \bra{1- \frac{\rho^2}{r_c^2}} \exp \bra{- \vx^2/r_c^2}.
 \end{align}
 
\begin{figure}
 \centering
 \begin{subfigure}{0.5\textwidth}
  \includegraphics[width=\textwidth]{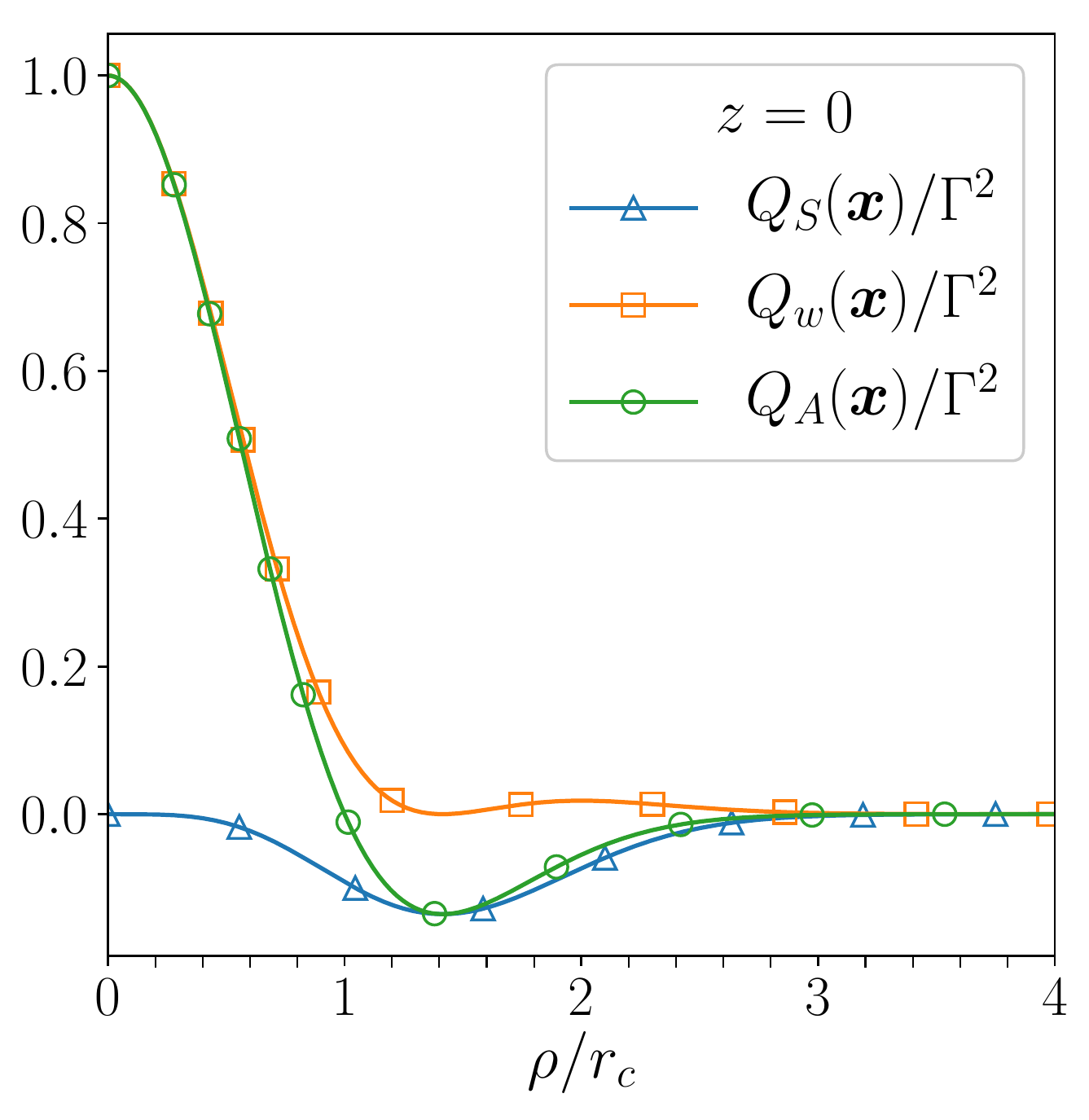}
  \caption{\label{fig: isolated vortex a}} 
 \end{subfigure}%
 \begin{subfigure}{0.5\textwidth}
  \includegraphics[width=\textwidth]{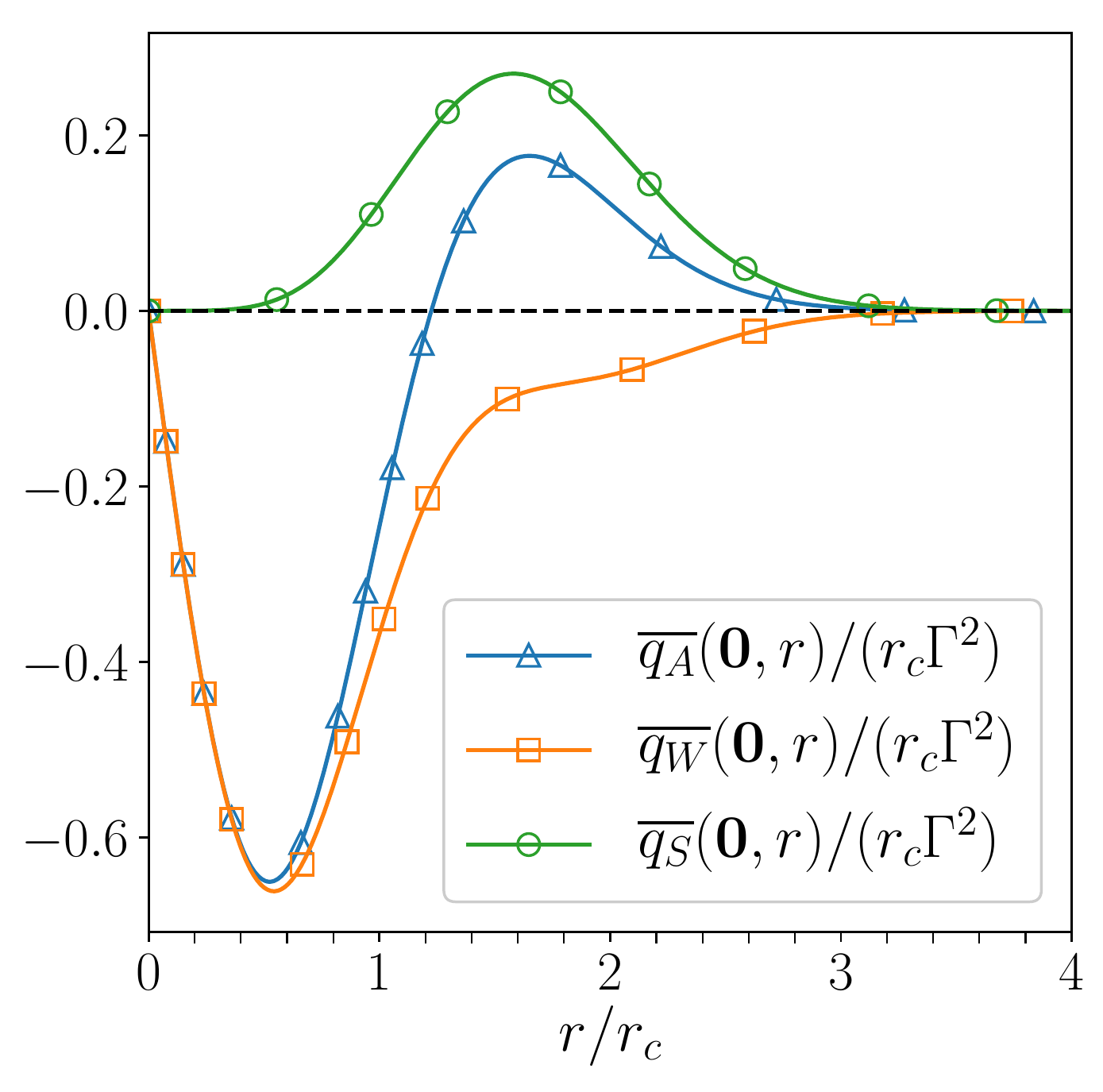}
  \caption{\label{fig: isolated vortex b}}
 \end{subfigure}%
 \caption{\label{fig: isolated vortex} 
 \small (\textit{a}) The radial dependence of $Q_S$ , $Q_W$ and $Q_A$ for the Townsend eddy in the $z=0$ plane.
  (\textit{b}) The contributions to the pressure at the origin $\vx=0$ from \emph{spherical} shells of given radius. 
  Note that $\overline{q_A} = \overline{q_S}+\overline{q_W}$.
  In total, the contributions to the pressure from the strain shell cancel out half of the contributions from the vortex core.
  }
\end{figure}

The central vortex has a characteristic width of $\sqrt{2}  r_c $ and is associated with a shell of enhanced strain localized around $|\vx|= \sqrt{2} r_c$ of the same characteristic width.
A slice through the $z=0$ plane illustrating this can be seen in \cref{fig: isolated vortex a}.
The peak intensity of the shell scales with strength of the central vortex.
Furthermore, the radius of the shell is clearly identifiable from the profile of $Q_A$.
Due to the overlap of the shell with the vortex, the region where $Q_A>0$ is smaller and has steeper gradients.
The attenuation might be small, but it has a significant impact on the resulting contributions to the pressure. 

As an example, consider the pressure at the origin (the point of largest enstrophy). 
The contributions to it from a given distance is shown in \cref{fig: isolated vortex b}.
The total contributions are again decomposed into strain and vorticity components according to $\overline{q_S} = -2 r \overline{Q_S}$ and $\overline{q_W} = -2 r \overline{Q_W}$, with $\overline{q_A} = \overline{q_W}+\overline{q_S}$.
It is clear that as far as the pressure is concerned the effective size of the vortex is much smaller, i.e. the $\overline{q_A}<0$ region is much smaller than the $\overline{q_W}<0$ region.
The amplitude of the pressure itself is also strongly reduced -- computing the integrals over the individual curves shows that the dissipation shell cancels out half of the vorticity contributions.

This simple model suggests that pure kinematics may be sufficient to explain at least to some extent the proximity of extreme vorticity and strain structures in real turbulent fields.
It also highlights how this proximity may attenuate the non-locality of the resulting pressure.
Because the calculations above are purely kinematic, they are quite robust and similar qualitative conclusions apply 
for other model vortices such as the Burgers and Rankin vortices as well.

\section{Distributions of velocity-gradient invariants}
\label{app: PDFs}

Here we describe the PDFs of the $Q_A$, $Q_S$ and $Q_W$ fields
and note how they can be used to derive high-order unconditional two-point statistics of $Q_A$ from the  conditional averages discussed in \cref{sec: cond vel gdt}.

\begin{figure}
\begin{subfigure}{0.33\textwidth}
	 \includegraphics[width=\textwidth]{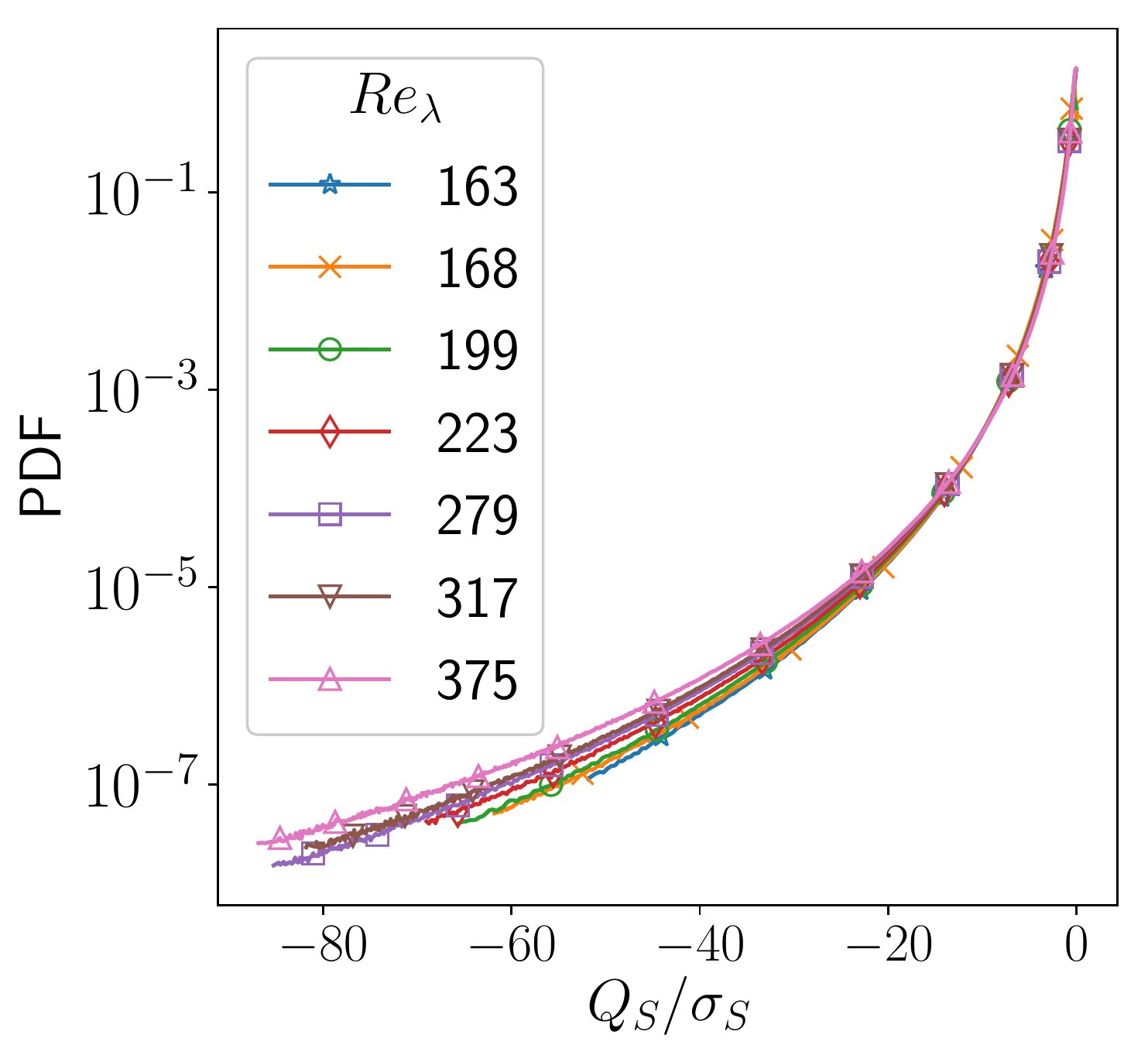}
	 \caption{}
 \end{subfigure}%
 \begin{subfigure}{0.33\textwidth}
	 \includegraphics[width=\textwidth]{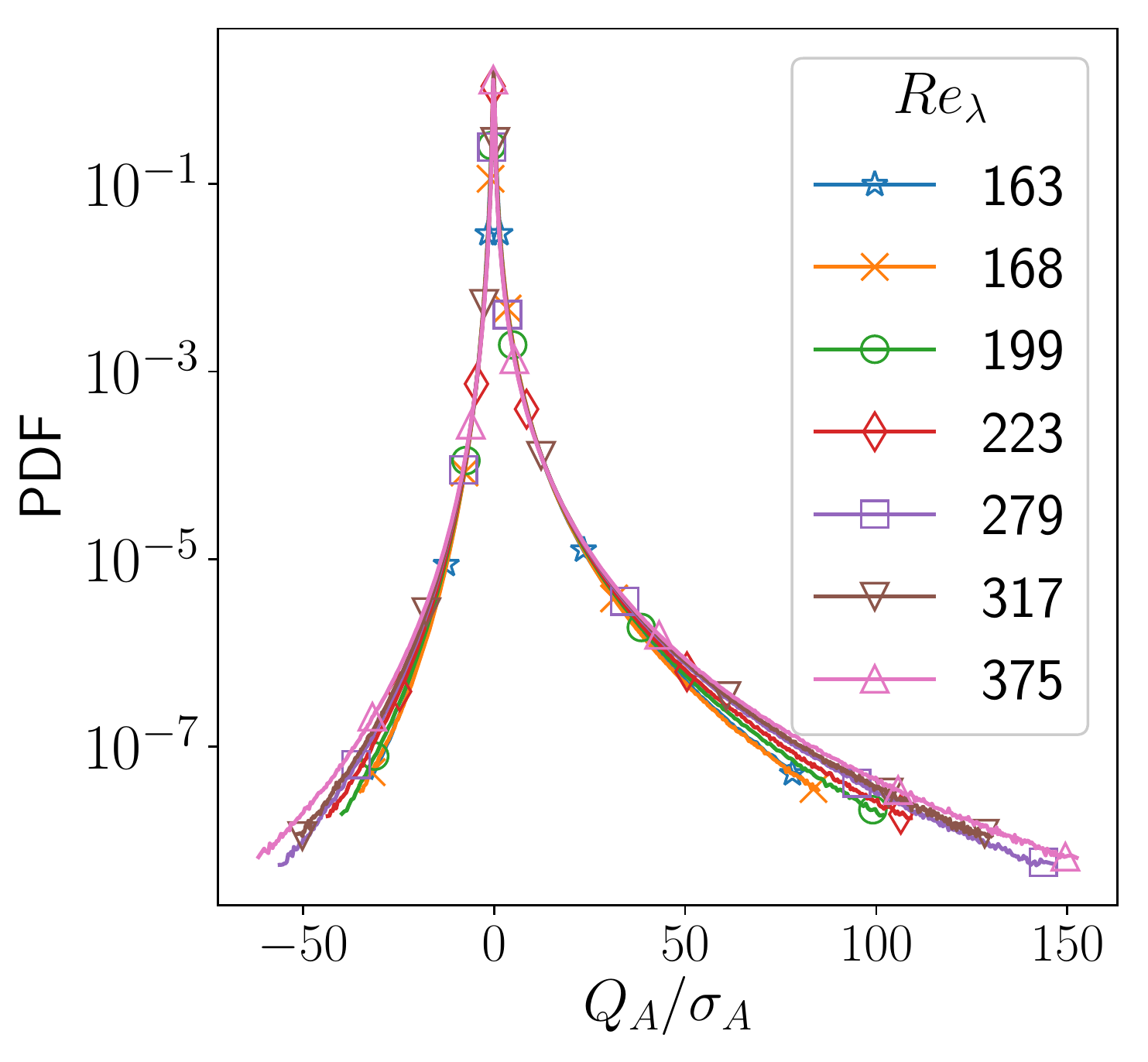}
	 \caption{}
 \end{subfigure}%
 \begin{subfigure}{0.33\textwidth}
	 \includegraphics[width=\textwidth]{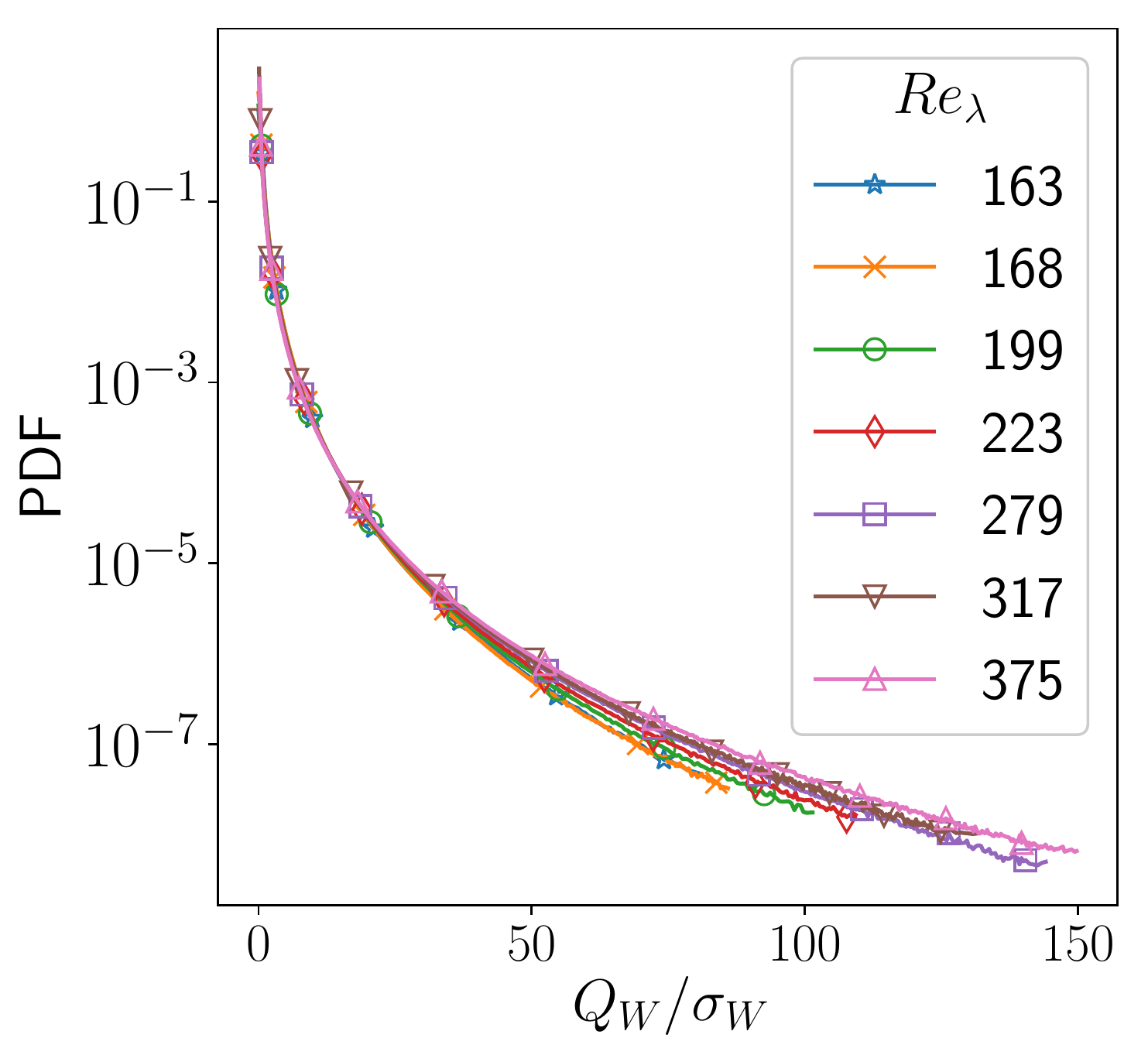}
	 \caption{}
 \end{subfigure}%
\caption {
\label{fig: pdf}
\small Probability density functions of (\textit{a}) $Q_S$, (\textit{b}) $Q_A$, and (\textit{c}) $Q_W$ rescaled by the respective standard deviations.
}
 \end{figure}

\Cref{fig: pdf} shows the respective PDFs for all analysed simulations. 
They are normalized by the respective standard deviations to highlight the extent of the extreme fluctuations.
This significantly reduces the Reynolds number dependence of the PDFs for intermediate and extreme values.
It should be pointed out that \cref{fig: pdf} illustrates events which have occurred more than $5\times10^5$ times over the total averaging volume and time. 
This truncation reduces the numerical uncertainty due to under-sampling in the tails of the PDFs.

The computed PDFs are in agreement with previous studies \citep[e.g.][]{Moisy2004, Yeung2012}. 
For lower Reynolds numbers the $Q_W$ field has heavier tails than $Q_S$, which is reflected in the positive skewness of $Q_A$. 
In other words, extreme enstrophy-dominated regions are more likely than their strain counterparts.
Furthermore, with increasing Reynolds number the frequency of extreme events slowly grows as expected,
 and so does the asymmetry between strain- and enstrophy-dominated regions.
These observations can be briefly quantified by the standard deviations of the fields and the skewness of $Q_A$ listed in
\cref{table: moments}.

\begin{table}
\centering 
\makebox[\textwidth]{
 \begin{tabular}{|c|c|c|c|c|c|}
 
 \hline
  \# & $\Rey_\lambda$ &  $\sigma_W/\avg{Q_W}$ & $\sigma_S/\avg{Q_W}$& $\sigma_A/\avg{Q_W}$ & skew$(Q_A)$
  \\
   \hline                       
   1 & 375  & 2.55 & 1.69 & 2.16 & 14.5
   \\  
   2 & 317  & 2.48 & 1.63 & 2.11 & 13.2
   \\
   3 & 279  & 2.43 & 1.60 & 2.07 & 12.5 
   \\                             
   4 & 223  & 2.31 & 1.52 & 1.98 & 11.1
   \\                             
   5 & 199  & 2.24 & 1.48 & 1.93 & 9.9
   \\                             
   6 & 168  & 2.12 & 1.41 & 1.83 & 8.8
   \\                             
   7 & 163  & 2.14 & 1.42 & 1.85 & 8.7
   \\
   \hline
 \end{tabular}
 }
 \caption{\label{table: moments} 
 \small Standard deviations of the $Q_W$, $Q_S$ and $Q_A$ fields given in units of $\avg{Q_W}$ and skewness of $Q_A$ for all examined simulations.
 }
\end{table}

Moreover, the computed PDFs are fitted well by a stretched exponential of the form $\mathcal{P}(F) = a_F \exp\bra{-b_F \abs{F}^{c_F}}$, 
adopted from \cite	{Hosokawa1991, Meneveau1991, Bershadskii1993}, namely 

\begin{align}
\mathcal{P}(Q_S) &= a_S \exp\bra{-b_S \bra{-Q_S}^{c_S}},\nonumber \\
\mathcal{P}(Q_W) &= a_W \exp\bra{-b_W \bra{Q_W}^{c_W}}, 
\label{eq: pdf fit}                      \\ 
\nonumber
\mathcal{P}(Q_A) &= a_A \times \begin{cases}
			\exp\bra{-b_A^+ \bra{Q_A}^{c_A^+}} & \textrm{ for } Q_A \geq 0\\
			\exp\bra{-b_A^- \bra{-Q_A}^{c_A^-}} & \textrm{ for }  Q_A < 0.\\
                       \end{cases} 
\end{align}
For the case of $Q_A$ the positive and negative branches of the PDF are fitted separately in order to account for the significant skewness of $Q_A$.
The fitting is performed using the Levenberg--Marquardt algorithm implemented by \cite{Newville2014}.
The details of the best-fit parameters are shown in \cref{table: pdf fit}. 
In agreement with \cite{Donzis2008} the coefficients $c_A^+ = c_A^- = c_S=c_W=0.25$ are kept fixed.
When included in the fitting, they vary from $0.2$ to $0.4$ with decreasing Reynolds number, but do not improve significantly the quality of the fits.
Because the far tails of the PDFs are not considered, the double stretched exponential proposed by \cite{Donzis2008} is found to be unnecessary.

Finally, we note that the fast decay of the PDFs has an important consequence for the correlation functions of $Q_A$, $Q_S$ and $Q_W$ .
Consider the cross-correlations $\mathcal{C}_{SW}$ as an example.
This can be expressed as 
\begin{align}
\mathcal{C}_{SW}(r)
		       &=\frac{1}{\sigma_S \sigma_W}\int \cavg{Q_S(r)}{Q_W=w} w \; \mathrm{PDF}_{Q_W} (w)\; \mathrm{d}w - \frac{ \avg{Q_S}^2}{\sigma_S \sigma_W},
\end{align}
using $\avg{Q_S} = \avg{Q_W}$.
Because the PDFs decay quasi-exponentially, extreme events have significantly reduced impact on the two-point correlation
 and on the resulting correlation length.

\begin{table}
\centering 
\makebox[\textwidth]{
 \begin{tabular}{|c|c|c|c|c|c|c|c|c|}
 \hline
  \# & $\Rey_\lambda$ & $a_W$ & $b_W$ & $a_S$ & $b_S$ & $a_A$ & $b_A^+$ & $b_A^-$  
  \\                                                                    
   \hline                                                               
   1 & 375            & 35    & 6.5   & 319   & 7.7   & 33    & 6.5     & 7.9   
   \\                                                                 
   2 & 317            & 57    & 6.7   & 534   & 8.0   & 51    & 6.7     & 8.2   
   \\                                                                   
   3 & 279            & 64    & 6.8   & 638   & 8.1   & 58    & 6.8     & 8.3   
   \\                                                                  
   4 & 223            & 103   & 7.0   & 861   & 8.3   & 94    & 7.0     & 8.6   
   \\                                                                  
   5 & 199            & 79    & 7.0   & 554   & 8.2   & 83    & 7.1     & 8.6   
   \\                                                                  
   6 & 168            & 164   & 7.3   & 802   & 8.3   & 162   & 7.4     & 9.0  
   \\                                                                  
   7 & 163            & 87    & 7.1   & 544   & 8.2   & 90    & 7.2     & 8.7   
   \\
   \hline
 \end{tabular}
 }
 \caption{\label{table: pdf fit}
 \small Best fit coefficients for the PDFs of $Q_S$, $Q_W$ and $Q_A$ for
 the functional forms given in \cref{eq: pdf fit}. The coefficients $c_F$ are kept constant at $0.25$.
 }
\end{table}

\section{Numerical computation of angle-averaged fields}
\label{app: angl ave}

For the purpose of reproducibility of the results we describe in brief the numerical recipe used to compute the analysed angle-averaged fields.
The non-trivial issue here is that of computing an average over a spherical shell for data given on a Cartesian grid.
To increase computational efficiency this was implemented as a convolution with a masking kernel $G(r)$.
This takes advantage of the convolution theorem for Fourier transforms and the efficiency of the fast Fourier transform algorithms.
Thus, the field $\overline{Q_A}(r)$ is computed using the formula
\begin{align}
 \overline{Q_A}(r) = \mathcal{F}^{-1}\bra{\widehat{G}(r) \widehat{Q}_A},
 \label{appeq: method}
\end{align}
where $\mathcal{F}^{-1}$ denotes the inverse Fourier transform operator and $\; \widehat{}\;$ denotes fields in Fourier space.

Given the finite grid resolution, we considered a kernel associated with spherical shells of finite thickness $\Delta$ (equal to the grid spacing). 
Analytically, the kernel and its Fourier transform (also known as a transfer function) can be written as 
\begin{align}
G (\vx, r, \Delta ) &= \frac{1}{V} \bra{\Theta\bra{|\vx| - r+\frac{\Delta}{2}} - \Theta\bra{|\vx|-r-\frac{\Delta}{2}}},
\label{appeq: G anal}\\
 \widehat{G}(\vk, r, \Delta) &=  \frac{8 \pi}{V|\vk|^3}  \left( \sin\bra{ |\vk| \Delta} \cos\bra{|\vk| r } \right. \nonumber \\
 &\;\;\;\;\;\;\;\;\;\;\;\;\;+ |\vk| r \sin\bra{ |\vk| \Delta}\sin\bra{|\vk| r} \nonumber \\
 &\;\;\;\;\;\;\;\;\;\;\;\;\;- \left.  |\vk| \Delta \cos\bra{ |\vk| \Delta} \cos\bra{|\vk| r}   \right),
\end{align}
 where $r$ is the shell radius, $\Theta$ is the Heaviside step function and $V$ is the volume of the thick spherical shell.
 
The most direct way to proceed is to discretize $\widehat{G}$ on a Cartesian grid in Fourier space. 
However, the discontinuities in \cref{appeq: G anal} mean that the discrete inverse Fourier transform of this field
is strongly affected by Gibbs oscillations and overshooting. 
As a result, it does not represent a radially compact kernel in real space, even for fairly large shell radii. 

To address this we discretize $G$ on a Cartesian grid in real space and take its Fourier transform numerically.
The discretization is performed by associating a grid cell with a particular spherical shell, if its centre is within the shell. In order to maintain the correct volume averages, e.g. $\avg{Q_A}=0$, we use a discretized value for $V$ which is consistent with the shell definition.

The finite grid spacing and the discontinuity in the analytical kernel impose uncertainties associated with this method as well.
In particular, the representative radius of a shell is not located half-way between its edges.
This uncertainty is incorporated in \cref{fig: thresh rad}.
It is largest at the smallest radii, but is always less than $\Delta/2$.
Because of it, when computing the pressure contribution profiles $\cavg{\overline{q_A}(r)}{Q_A}$, the weighting of the Poisson kernel has to be included on a cell-by-cell basis. 
This ensures that the resultant conditional pressure average $\cavg{p(\vx)}{Q_A(\vx)}$ computed from the angle-averaged profiles is consistent with the one computed directly.

\bibliographystyle{jfm}
\bibliography{master}

\end{document}